\documentclass[fleqn,10pt]{wlscirep}

\usepackage{graphicx,epsfig}

\usepackage{times}
\usepackage{graphics,dcolumn,bm,fleqn,float}
\usepackage{amssymb,amsmath,multirow,rotate,color}
\usepackage{comment} 
\usepackage{cite} 
\usepackage{stackengine}
\usepackage[normalem]{ulem}
\DeclareMathAlphabet{\altmathcal}{OMS}{cmsy}{m}{n}

\graphicspath{{./figures/}}

\newcommand{\vc}[1]{{\color{violet} #1}}
\title{Resilience and elasticity of co-evolving information ecosystems}

\author[1]{Mar\'ia J.~Palazzi}
\author[1,2]{Albert Sol\'e-Ribalta}
\author[3]{Violeta Calleja-Solanas}
\author[3]{Sandro Meloni}
\author[4]{Carlos A. Plata}
\author[4]{Samir Suweis}
\author[1]{Javier Borge-Holthoefer}
\affil[1]{Internet Interdisciplinary Institute (IN3), Universitat Oberta de Catalunya, Barcelona, Catalonia, Spain}
\affil[2]{URPP Social Networks, University of Zurich, Zurich, Switzerland}
\affil[3]{IFISC, Institute for Cross-Disciplinary Physics and Complex Systems (CSIC-UIB), 07122, Palma de Mallorca, Spain}
\affil[4]{Dipartimento di Fisica e Astronomia G. Galilei, Universit\`a di Padova, Via Marzolo 8, Padova, 35131, Italy}

 \begin{abstract}
Human perceptual and cognitive abilities are limited resources\cite{simon1972theories,kahneman1973attention,posner1982cumulative}. Today, in the age of cheap information --cheap to produce, to manipulate, to disseminate--, this cognitive bottleneck translates into hypercompetition\cite{chadwick2011political} for visibility among actors\cite{gonccalves2011modeling,weng2012competition,gleeson2014competition,gleeson2016effects} (individuals, institutions, etc). The same social communication incentive --visibility-- pushes actors to mutualistically interact with specific memes\cite{dawkins2016selfish}, seeking the virality of their messages. In turn, contents are driven by selective pressure, {\it i.e.}~the chances to persist and reach widely are tightly subject to changes in the communication environment.
In spite of all this complexity, here we show that the underlying architecture of the users-memes interaction in information ecosystems, apparently chaotic and noisy, actually evolves towards emergent patterns, reminiscent of those found in natural ecosystems\cite{bascompte2007plant,stouffer2011compartmentalization}. 
In particular we show, through the analysis of empirical, large data streams, that communication networks are structurally elastic, {\it i.e.}~fluctuating from modular to nested architecture as a response to environmental perturbations ({\it e.g.}~extraordinary events)\cite{borge2017emergence}. We then propose an ecology-inspired modelling framework\cite{suweis2013emergence}, bringing to light the precise mechanisms causing the observed dynamical reorganisation. Finally, from numerical simulations, the model predicts --and the data confirm-- that the users' struggle for visibility induces a re-equilibration of the network towards a very constrained organisation: the emergence of self-similar nested arrangements.
\end{abstract}

\begin{document}

\maketitle

\section*{Introduction}
Our current experience of the accelerated stream of digital content\cite{lorenz2019accelerating} has exposed, in full range, the tight bio-cognitive limitations that we are subject to\cite{simon1972theories,kahneman1973attention,posner1982cumulative}. Yet, their finiteness had not, in general, arisen in quotidian communication processes: not in the pre-industrial age, where physical (face to face) or low-bandwidth interaction governed the slow change of public opinion; and neither during the predominance of mass media, when the exposure to an oligopolistic media environment put little pressure to the attentional resources of the audience. In both cases, the public sphere was hierarchically structured and framed by the operations of few actors on a rather slow time scale. Contrarily, the paradigm of online communication is characterised by the fragmentation of the public sphere\cite{bruns2015habermas}, in which elite and non-elite actors behave like information sources and receivers on the virtual stage. Only in this new scenario, attention, memory and processing time suddenly become critical assets to compete for\cite{gonccalves2011modeling,weng2012competition,gleeson2014competition,gleeson2016effects}: their scarcity has been exposed.

Complementary to direct competition (among actors), interaction with other units in the system is often mutualistic. For the same reason that two actors compete with each other, they establish cooperative relationships with the memes (keywords, hashtags). These ``information chunks'' may --if correctly chosen-- optimally spread information and consolidate the visibility they strive for. Hence, for example, the (ab)use of hyper-emotional language that we suffer in nowadays politics, as an arms race to impact optimisation. 

Of course, the choice of a meme is context-dependent (``past performance is no guarantee of future results''), and thus the interactions between actors and memes are co-evolving and extremely sensitive to changes in the communication environment --breaking news, fads and rumours, celebrity gatherings, etc--. In turn, changes in the surrounding conditions tend to be ephemeral although frequent, in the more open and fluid access to many digital sources.

Under the light of these four drivers --competition, mutualism, co-evolution, environment--, online communication systems become a special case of mutualistic ecosystems\cite{bascompte2007plant}. Our failure to realise this in the past (despite some clues\cite{gonccalves2011modeling,weng2012competition,gleeson2014competition,gleeson2016effects,borge2017emergence}) is perhaps due to the extreme difference between the time scales that dominate the two relevant components at stake: the slow-changing social contacts (user-user) network, and the accelerated dynamics through which information is created and spread\cite{lorenz2019accelerating} --the former can be safely considered static if compared to the latter\cite{vespignani2011modelling}, completely missing the co-evolutive aspect of the system--. However, this picture changes dramatically if the focus is shifted from the relatively stable peer-to-peer network to the fluid information network, that is, {\it ad hoc} groups of users, which loosely gather around and engage in shared memes \cite{bruns2011use}, operating in a hyper-competitive environment\cite{chadwick2011political}.

This revealing picture opens new promising possibilities to analyse and model online social networks, if we consider that Ecology is rich in theoretical frameworks where the co-evolutionary interplay between structure and dynamics is studied\cite{suweis2013emergence,guimaraes2017indirect,cai2018dynamic}. Moreover, while testing these theories empirically in natural ecosystems is difficult --mainly because of the resource-intensive demands to collect data\cite{pilosof2017multilayer}--, digital streams from social interactions are abundant on several spatial and temporal scales, and precise knowledge about the environmental (external) conditions --related to specific information flows-- can also be collected.

The first problem to address under this {\it information ecosystems} framework is the network's structural volatility, which is coupled to the fluctuating nature of the environment. Online communication is heavily driven by the events surrounding it, which constantly trigger attention shifts that modify the behaviour of otherwise loosely linked assemblages of individuals and groups\cite{chadwick2011political}. It is precisely this hectic, information-dense environment that dictates the emergence and fall of ephemeral synchronised attention episodes, which translate in fast structural changes. 

Here, we provide evidence that information ecosystems exhibit a remarkable structural elasticity to environmental changes. To do so, we first report on theory-free, empirical observations of the characteristic dynamical re-organisation in communication networks, as they react to environmental ``shocks''. Analysing the response of the Twitter ecosystem to different types of external events, we quantify how collective attention episodes reshape the user-hashtag information network, from a modular\cite{newman2004finding,fortunato2010community} to a nested\cite{payrato2019breaking,bascompte2003nested} architecture, and back. The emergence of these structural signatures is, remarkably, consistent across different topics and time scales. 
Next, we propose a theoretical framework that explains the emergence of the patterns observed in real data streams, as a result of an adaptive mechanism. The model builds on the idea that the user-meme network structure is effectively driven by an optimisation process\cite{suweis2013emergence}, aiming at the maximisation of visibility, and that the nature of the user-meme interactions is mutualistic, i.e., beneficial for both. Furthermore, through our modelling framework we predict that the users' struggle for visibility in any context facilitates the emergence of nested self-similar arrangements: either mesoscale (in-block) nestedness\cite{lewinsohn2006structure,sole2018revealing} during the compartmentalised stages, or macroscale nestedness in exceptional global attention episodes. We finally show that these predictions are supported by the data.


\section*{Results}

\subsection*{Structural elasticity in information systems}
Biased as it may be\cite{gonzalez2014assessing}, Twitter is without a doubt a sensitive platform that mirrors, practically without delay, exogenous events occurring in offline environments. In this sense, Twitter data constitute a rich stream, providing a public and machine-readable vision of the non-virtual world.

Despite the highly fluctuating nature of this endless communication stream, some reliable patterns emerge from its apparently hectic activity. We analyse these streams in a longitudinal manner, as a series of snapshots from time-resolved activity. Each slice is represented as a bipartite network with a fixed number of most active users ($N_{U} = 2000$), and the corresponding hashtags $N_{H}$ created and/or cited by these users, see Methods Summary below.
Such sequence of networks is studied monitoring different structural arrangements that are relevant to the dynamical stage in which the system is. For now, we focus on two of them: modularity\cite{newman2004finding,fortunato2010community} ($Q$) and nestedness\cite{patterson1986patterson,atmar1993measureorder,bascompte2003nested} ($\altmathcal{N}$), see the Methods Summary. High levels of modularity correspond to a fragmented attention scenario, and can be considered as the {\it resting state} of the system. In this stage, users mostly focus on their own topics of interest, {\it i.e.}~a certain subset of memes, facilitating the emergence of identifiable blocks. High values of nestedness, on the other hand, reflect an extraordinary (and, thus, ephemeral) stage in which the system self-organises to attend one or few topics. In these cases, the discussion revolves around a small set of generalist memes (hashtags used virtually by everybody) and users (highly active individuals participating in many facets of the discussion).

\begin{figure}[h]
\centering
\topinset{\bfseries (a)}{\includegraphics[width=.47\textwidth]{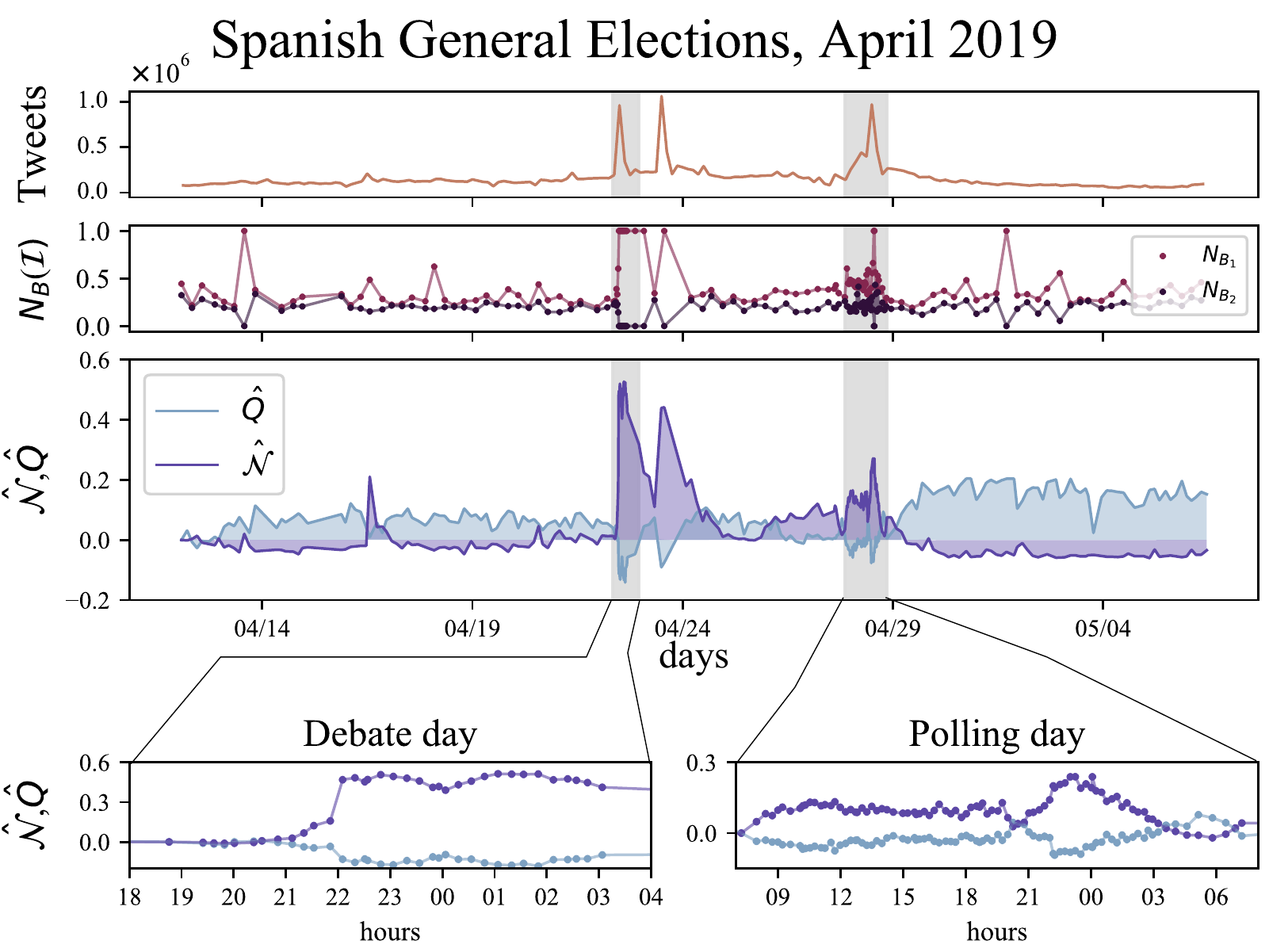}}{0.075in}{1.4in}
\topinset{\bfseries (b)}{\includegraphics[width=.47\textwidth]{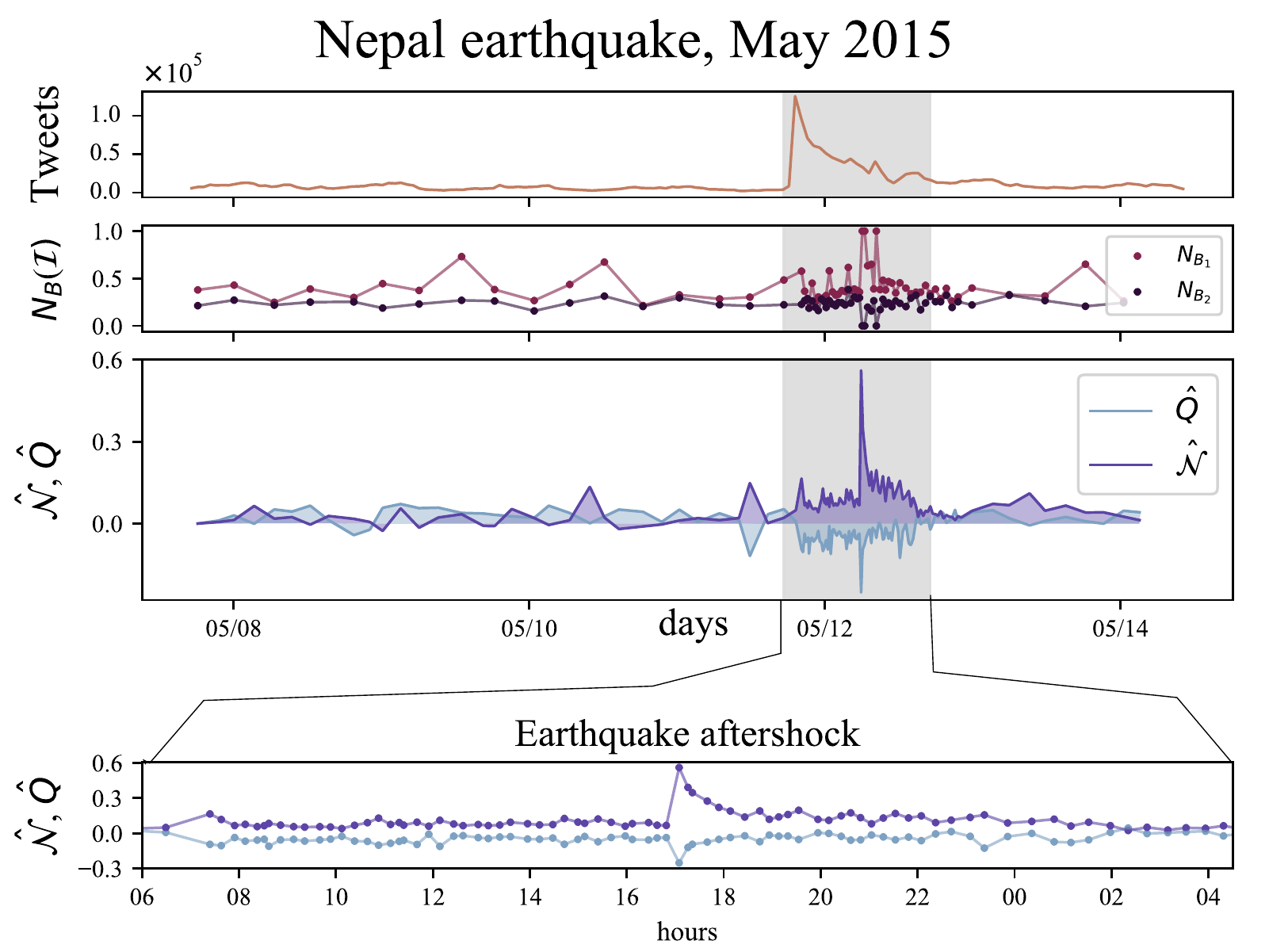}}{0.075in}{1.4in}
\caption{{\bf Structural measures over time for two datasets.} Here we show Twitter streams covering two different topics, i.e. Spanish general election of 2018 (panel \textbf{a}) and 2015 Nepal earthquake (panel \textbf{b}). Spanning different time ranges and attracting varying levels of attention (see tweet volume in top panels), the information ecosystems self-organise in similar ways: a block organisation dominates the system (positive modularity $\hat{Q}$), reflecting the separate interests of users, until external events induce large-scale attention shifts, which rearrange completely the network connectivity towards a nested architecture (high $\hat{\altmathcal{N}}$). Note that, despite the predictable (Spain) {\it vs.} unpredictable (Nepal) nature of each stream, structural properties of the user-meme interaction networks ($\hat{Q}$ with $\hat{\altmathcal{N}}$) are anti-correlated\cite{borge2017emergence}.
For a closer view, we highlight specific time windows in each dataset with some identifiable events happening in them (lower panels). In each plot, measures of modularity and nestedness are shifted from their initial values (that is, $\hat{Q}(t) = Q(t)-Q_{0}$ and $\hat{\altmathcal{N}}(t) = \altmathcal{N}(t) - \altmathcal{N}_{0}$, respectively).
The panels corresponding to $N_{B}(\altmathcal{I})$ highlight the nested self-similar arrangements at different scales, which is discussed later on.}
\label{fig:empirical}
\end{figure}

Figure~\ref{fig:empirical} presents the evolution of $Q$ and $\altmathcal{N}$ on two different portions of Twitter activity (several more are shown in section 3.1 of the Supplementary Material, with similar insights). For example, Fig.~\ref{fig:empirical}a corresponds to a period of over 45 days around the local elections in Spain (April-May 2019) (see section 1.1 of the Supplementary Material for details). For this dataset, the evolution of $Q$ and $\altmathcal{N}$ shows a perfectly anti-correlated behaviour. Such behaviour can be explained by the mutual structural constraints that these two arrangements impose on each other\cite{palazzi2019antagonistic}. Remarkably, however, the significant growth of nestedness is not merely due to the fluctuating character of the system, but, on the contrary, tightly linked to external events: see for instance the sudden changes in the structure on specific dates, shadowed in grey  in the figure (debate and polling day, respectively). These extraordinary events are accompanied, unsurprisingly, by an increased volume of messages (top panel). The figure, at the scale of days, is complemented with high-resolution monitoring of portions of these exceptional events (bottom panel), which confirm the general anti-correlated trend. Finally, the most outstanding feature highlighted by the figure is the elasticity of the network: no matter how abrupt and large the excursion to a nested arrangement is, the system bounces back to its ``ground'' --predominantly modular-- state soon after, when the interest in the breaking news fades out. The observed elasticity can be considered as an aspect of the network's structural resilience. System resilience or stability is defined in different ways in ecology and environmental science\cite{holling1996engineering,folke2010resilience,ives2007stability}, but can generally be thought as the ability of the system to recover the original system's state after a perturbation of the model state variables\cite{suweis2015localization,arnoldi2016resilience} or parameters\cite{rohr2014structural,grilli2017feasibility}. Specifically, in the case of structural elasticity, the system state is not given by the nodes' configuration ({\it e.g.}~the abundance of each species), but by the overall network architecture ({\it i.e.}~modular, nested), which is perturbed by the external event. 

This behaviour is stable across different types of event. Figure~\ref{fig:empirical}b shows an equivalent behaviour for a completely different event. In this case, the dataset comprises the reaction after the Nepal earthquake in 2015\cite{zubiaga2018longitudinal}, including a major aftershock on May 12th. Unlike a political debate or an election date, this example is inherently unexpected and unpredictable --an important fact, attending the taxonomy of collective attention described in Lehmann {\it et al.}\cite{lehmann2012dynamical}. As in Fig.~\ref{fig:empirical}a, the coarse grain scale of days and weeks in Fig.~\ref{fig:empirical}b is complemented with high-resolution monitoring a portion of exceptional events. See Supplementary Material for additional examples.


These analyses suggest that there is a tight logic underlying the structural fluctuations of the information network: the level of fragmentation of collective attention maps onto specific network arrangements, and is independent of the particular contents of the data stream. Online activity on different topics translates to comparable changes in the resulting patterns, no matter the semantics of the underlying discussion. The observed differences in the emergence, magnitude and persistence of structural changes are directly related to the predictability, intensity and duration of the exogenous events ({\it i.e.}~related to the environmental conditions), and therefore cannot be explained as intrinsic to the communication system itself. 
The question remains, however, how a networked system can fluctuate so fast between two states which have often been considered incompatible\cite{Thebault2010,fortuna2010nestedness,palazzi2019antagonistic}.

\subsection*{Theoretical Framework}
To understand the mechanisms that govern the observed elasticity, and, at the same time, to solve the puzzle around the network's nested-modular oscillations, we propose a model founded on the ecological drivers introduced above: competition, mutualism, co-evolution and environment. The model builds on the simple idea that the network architecture between users and memes is the result of several local optimisation processes, {\it i.e.} each individual's maximisation of visibility, and that such process operates on top of attentional dynamics. To do so, we generalise the ecological adaptive modelling proposed by Suweis {\it et al.}\cite{suweis2013emergence,cai2018dynamic}, in which the system's actors (plant and pollinator species) strive for larger individual abundance, rewiring their interactions accordingly.

\paragraph{The Model.}

The synthetic information network model comprises a total of $N$ interacting ``species'' or nodes ($N_{U}$ users and $N_{H}$ hashtags or memes), in which population dynamics --where population here quantifies the visibility of the users and/or of the memes-- is driven by interspecific mutualistic interactions, following a Lotka-Volterra dynamics with Holling-Type II functional response \cite{bastolla2009architecture,suweis2013emergence}. 

Each species has an associated niche\cite{Williams2000simple} which, in the context of an information ecosystem, represents their topical domain ({\it i.e.} the topic to which a user attends preferentially, and, conversely, the semantic space where a meme belongs to). For the sake of simplicity, each species' niche is represented as a Gaussian distribution with a given standard deviation $\sigma$\cite{cai2018dynamic}. 
Both users and memes niches are anchored around $T$ different points in the range $[0,1]$, to express different topic preferences (users), and semantic domain (memes). To model the inherent diversity of users and memes within their topic, their position over the line is perturbed by a small amount, randomly sampled from a uniform distribution.

\begin{figure}[h!]
\centering
	\includegraphics[width=.95\textwidth]{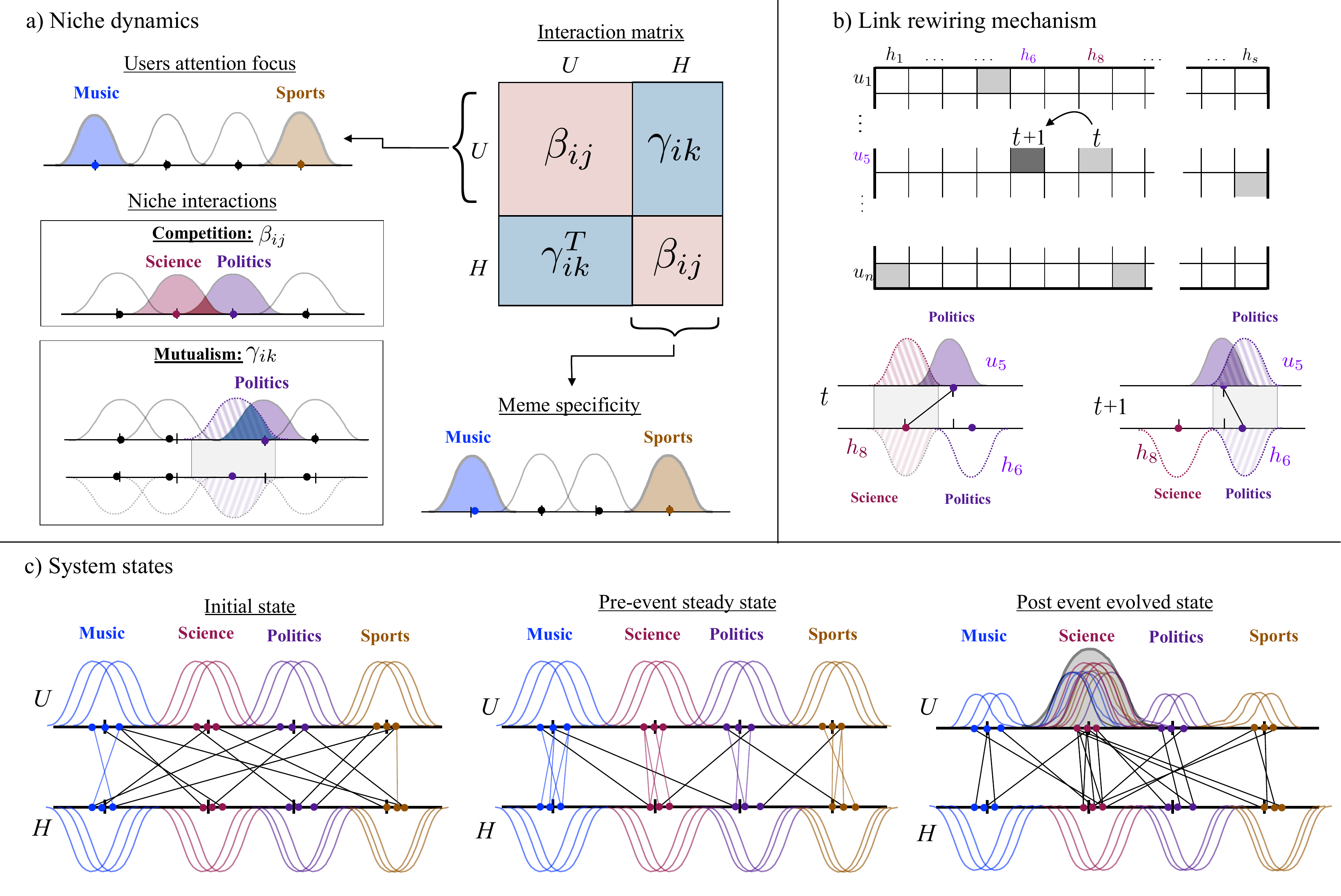}
\caption{{\bf Schematic representation of the visibility optimisation model.} \textbf{(a)} Users ($U$) and memes ($H$) are represented as points in the range $[0,1]$ with their niche as a Gaussian curve with standard deviation $\sigma$. Topics are modelled as clusters of users (memes), {\it i.e.}~$T=4$. \textbf{(b)} At each time-step, species rewire their connections trying to optimise their abundance (popularity). If the rewiring leads to a larger popularity the connection is kept, otherwise the change is reverted. \textbf{(c)} When an external event enters the system, user niches are temporarily focused on a specific topic (Science in the illustration) and the rewiring takes place. As the event fades out, all species return to their original niche.}
\label{fig:model}
\end{figure}

Competition occurs between species of the same class (or guild), whereas mutualistic interactions couple the dynamics of abundance of users and memes. Following the proposal of Cai~{\it et al.}\cite{cai2018dynamic}, the strength of the competitive interactions between a pair of users (memes) is tuned by a fixed parameter ($\Omega_{c}$) scaled by a quantity that depends on the niche overlap $G_{ij}$ between them. Similarly, the strength of the mutualistic interactions between a pair user-meme results from a fixed parameter ($\Omega_{m}$) scaled by the niche overlap between the pair user-meme --{\it i.e.}~the similarity between the user's topic preference and the adequacy of the meme within this topic--, and constrained to the existence of a link between them. Figure~\ref{fig:model}a summarises the ingredients of the model. We note that, in contrast to natural ecosystems, memes are an infinite resource --which explains why user-user competition does not grow with the amount of shared memes.

On the dynamical side, each user attempts to change its mutualistic partners (memes) in order to maximise the benefit obtained from their use (see Supplementary Material). This optimisation principle may then be interpreted within an adaptive framework, in which users incrementally enhance their visibility by choosing the appropriate memes, and memes are created so as to maximise their diffusive capacity, see Figure~\ref{fig:model}b. In summary, both classes optimise the efficiency of resource usage, decreasing their chances of becoming extinct due to stochastic perturbations\cite{gleeson2014competition}. Within the model, this translates into reiterative rewiring interactions of randomly drawn users so as to increase their visibility --``abundance'' in the ecological jargon.

Since our primary objective is to reproduce structural changes under the irruption of external events, the dynamical model includes as well a mechanism to introduce exogenous events in the environment. These can be understood as transitory shifts in the users' attentional niches, which are tantamount to (typically short-lived) changes in their interests (Figure~\ref{fig:model}c). In this altered environment, users temporarily engage with new kinds of hashtags, different from those they usually interact with (see the Methods Summary and Supplementary Material).

\begin{figure}[h!]
\centering
   \topinset{\bfseries (a)}{\includegraphics[width=.47\textwidth]{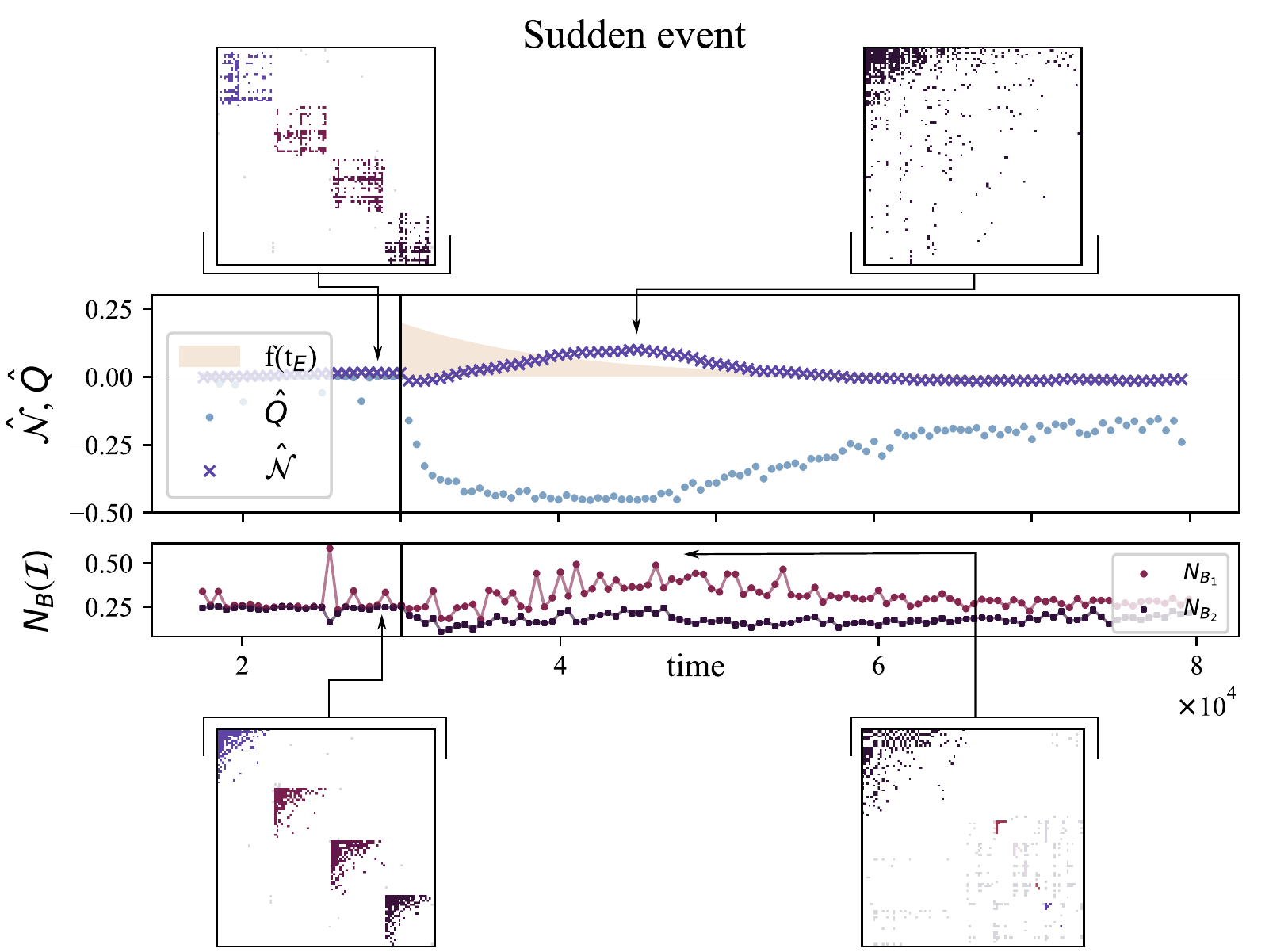}}{0.075in}{1.4in}
   \topinset{\bfseries (b)}{\includegraphics[width=.47\textwidth]{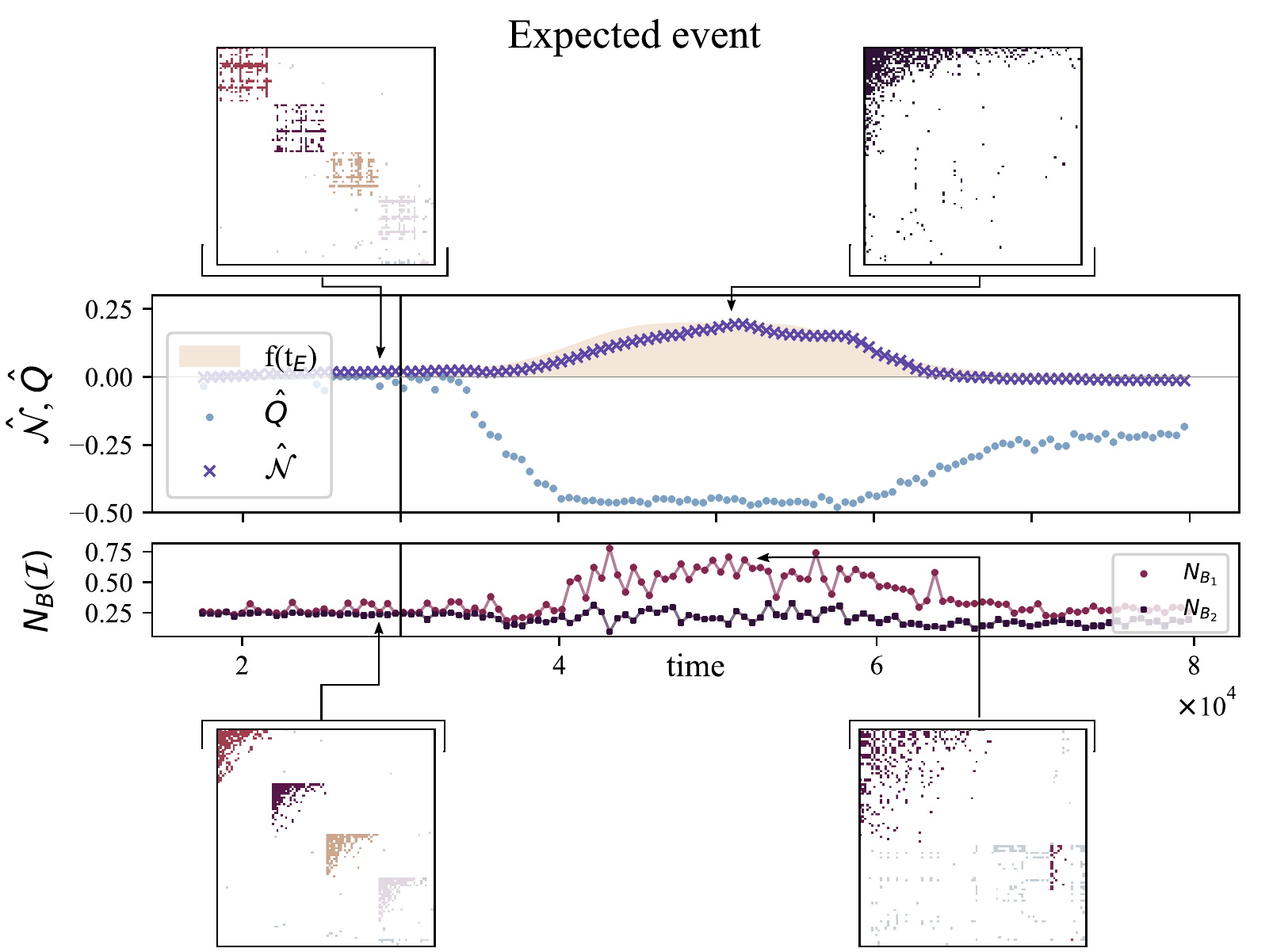}}{0.075in}{1.4in}
\caption{{\bf Structural evolution in the visibility optimisation model.} The figure corresponds to controlled numerical experiments at the stable stationary state, by holding fixed the number of species (100 users and 100 hashtags) and connectance ({\it i.e.} the fraction of non-zero interactions) $C=0.058$. The dynamics seeks to maximise individual species abundances by varying the network architecture. Initially, links between users and hashtags are laid at random. Users and memes are aligned with a number of predefined topics ($T=4$, as in Figure~\ref{fig:model}). Panel (a) mimics the arrival of an unexpected exogenous event, represented as a monotonically decreasing yellow shade starting at $t = 3 \times 10^{4}$. Panel (b), instead, models the increase, sustainment and decay of attention in programmed events (e.g. election day), represented also as a yellow shade.
In the absence of an exogenous event, and following the trend observed in empirical data, the model initially organises in a clear block structure. Once the external event enters the system, the network blurs its modular organisation (abruptly in panel (a); smoothly in panel (b)), and evolves towards a hierarchical, nested configuration. After the effects of the shock fade, the network slowly recovers its baseline modular configuration. The adjacency matrices surrounding the plots show the block and in-block nested structure of the bipartite network immediately before (top- and bottom-left panels, respectively) the onset of the perturbation, and the nested and in-block nested arrangement some time after (top- and bottom-right panels, respectively). In both panels, results correspond to an average over 10 realisations.
}
\label{fig:numerical}
\end{figure}

\paragraph{Structural evolution.} 
In an unperturbed simulated environment, the observed emergent structural arrangement mimics the prescribed organisation of niches in topical blocks. That is, a modular architecture arises from the random initial one, see Figure~\ref{fig:numerical} for $t < 3 \times 10^{4}$  (note that the plot is shifted by $Q_{0}$, {\it i.e.} modularity at time $2 \times 10^{3}$, once the system has stabilised its architecture). This is in line with the {\it resting state} observed in the datasets (Figure~\ref{fig:empirical}), where users are focused on their own topics of interest. It is important to underline that the emergence of a modular architecture is not an artefact of the model: users (memes) do not rewire by similarity reasons; it is the search for an improvement in their individual visibility that naturally drives to the consolidation of those new connections. Also note that, in empirical settings, the random initial stage is impossible to observe since the network already has a modular organisation from the very beginning.

An abrupt change in the environment --{\it e.g.}~breaking news-- totally alters this scenario. 
The system reacts almost immediately with a sharp decrease in $Q$, and an increase in the amount of nestedness in the system, Figure~\ref{fig:numerical}a for $3 \times 10^{4} < t \lesssim 4 \times 10^{4}$. A similar phenomenon occurs if the simulation refers to a predictable event, Figure~\ref{fig:numerical}b, except that the collapse of $Q$ is smoother, and the emergence of nestedness is slightly delayed. Indeed, in this situation we recover the results in Suweis {\it et al.}\cite{suweis2013emergence} --the emergence of global nestedness--, because the existence of attentional niches becomes irrelevant when all niches are equally centred, at least on the users' side. In this sense, our niche-based population dynamics is a generalisation of Suweis and co-authors' model. As the environmental shock fades out, the network architecture tends to recover the general layout present before the event was introduced, see $t \gg 4 \times 10^{4}$. The elasticity of empirical information ecosystems is thus replicated here, and explained as a consequence of the adaptation to contextual changes --while the species' local strategies remain constant. 

\paragraph{Nestedness reframed: multi-scale analysis.}
Beyond the examination of the evolution of $Q$ and $\altmathcal{N}$, we now take a closer look at the intra-modular organisation of connections during the fragmentary stage of the system ($t < 3 \times 10^{4}$). For visualisation purposes, the rows and columns of the adjacency matrices in the top-left part of Figures~\ref{fig:numerical}a and \ref{fig:numerical}b have been arranged to highlight the block structure that results from modularity optimisation. Additionally, rows and columns inside modules were sorted, in the bottom-left part, in order to highlight the possible nested structure within them\cite{flores2011statistical,flores2013multi}. Clear to the naked eye, each compartment presents an internal nested architecture. This is a natural consequence of the node-level visibility-maximisation strategy as it adapts to system-wide environmental conditions: as long as these conditions are stable around weakly connected topics, nestedness emerges in those relatively isolated subsystems. \vc{Moreover,} As soon as the boundaries across subsystems are blurred ($t > 3 \times 10^{4}$, top-right of Figures~\ref{fig:numerical}a and 3b), global nestedness prevails.

This subtle insight, which stems from the model, reframes the empirical findings presented above. Indeed, the information network is not swapping between two radically different architectures --often even antagonistic\cite{Thebault2010,palazzi2019antagonistic}--, but rather fluctuating across nested self-similar arrangements at different scales. To quantify them, $\altmathcal{N}$ is not a suitable tool, because it is designed to capture nestedness at the global scale only. For this reason, we resort to in-block nestedness $\altmathcal{I}$\cite{sole2018revealing,palazzi2019antagonistic,palazzi2019online}, which generalises $\altmathcal{N}$. On the one hand, when nestedness emerges at the global scale (one block, $B=1$), then we have that $\altmathcal{I} = \altmathcal{N}$. On the other hand, when the network presents several blocks ($B > 1$), each one arranged in a nested manner, then $\altmathcal{I} > \altmathcal{N}$.

It makes sense now to revisit the previous numerical and empirical results, now through the lens of in-block nestedness. Figure~\ref{fig:empirical} (second panels in (a) and (b)), and Figure~\ref{fig:numerical} (bottom panels in (a) and (b)) monitor the relative size of the largest ($N_{B_{1}}/N$) and second largest ($N_{B_{2}}/N$) nested blocks. In both empirical and numerical cases, we observe that nearly-perfect consensus is reached at different moments ($N_{B_{1}}/N \approx 1$) during the exogenous events, while a fragmented public sphere dominates most of the time. The relative size of the second largest nested block ($N_{B_{2}}/N$) allows for an easier interpretation of the level of consensus reached at each time.

Our framework allows to explain the puzzling transition between partial and global consensus. A fast re-organisation from modular (nested) to nested (modular) architectures seems paradoxical and hard to achieve. Nevertheless, the system can swiftly adapt to any state of collective attention through an intermediary arrangement that combines the structural signature of visibility maximisation with the existence of a fragmented public sphere. 

\section*{Discussion}
The transit from a secular hierarchical management of public information to a decentralised and fragmentary scenario calls for a new vision in which the relevant drivers are identified: competition for cognitive resources, mutualistic exploitation of content, co-evolution of users' and memes' visibility, and environmental conditions. So far, incursions in such ecological mindset have been sparse\cite{weng2012competition,gleeson2014competition,borge2017emergence,lorenz2019accelerating}. In this work, going beyond a simple metaphoric interpretation, we prove that an ecological framework --with explicit use of competitive and mutualistic interactions as drivers of information dynamics-- is a powerful tool to describe the evolution of information ecosystems. Indeed, although simple neutral models may account for emergent patterns in the popularity  distribution\cite{weng2012competition,gleeson2014competition}, we show that our non-neutral, niche-based population dynamics model can successfully explain the complex interplay between users-memes interactions, attentional niches and environmental shocks.

Our results open an ambitious research alley. In the shorter term, future efforts should attempt to mimic the microscopic dynamics of users and memes abundances before and after breaking events. These cannot be explained without including death-birth and invasion processes\cite{hui2017invasion}, which are in turn necessary to understand how influential users and viral contents emerge. Similarly, this initial proposal rules out ``cultural drift'' --the slower changes in the users' topical preferences--, which leads to persistent structures and shapes communication flows. 

Reaching further, the tradition in theoretical ecology aimed at understanding and preventing the collapse of ecosystems can be adopted to decipher how social media and information bubbles shape our thinking\cite{rahwan2019machine}, or, in the opposite direction to disrupt and break misinformation dynamics and polarisation. Related to this, we foresee as well a connection between the extensive research on stability and resilience in natural ecosystems, and their informational counterparts. In this sense, we are convinced that such interchange of techniques and models could be beneficial for theoretical ecology too as it will allow to test theories and methodologies in a more controlled, data-rich environment with faster time scale at play.

\section*{Methods Summary}

\paragraph{ Empirical and synthetic data.} 

\subparagraph{Matrix generation.}  For both synthetic and empirical cases, we represent a bipartite unweighted network as a $N_U \times N_H$ matrix $\mathbf{A}$, where rows and columns refer to users and hashtags, respectively. Elements therefore represent links in the bipartite network, {\it i.e.}~if the element $a_{uh}$ has a value of $1$, it represents that the user $u$ produced the hashtag $h$ at least once, otherwise $a_{uh}$ is set to $0$. 

For the generation of synthetic data, we set up a small network of 100 users and 100 hashtags, and the interactions between users and hashtags are laid at random with a fixed connectance of $C_0=0.058$. Then, for the empirical case, we build a sequence of snapshots by splitting the Twitter datasets into chunks containing the $2000$ most active unique users, while the number of hashtags is variable (depending on the amount produced by those 2000 users). In this way, for each snapshot, a rectangular binary presence-absence matrix is created. See Figure~S1 and surrounding text in the Supplementary Material for further details on the construction of networks.

\subparagraph{Structural measures.}
In this work, we explore the structural evolution of the network by means of three arrangements, one at the macroscale (nestedness\cite{patterson1986patterson,atmar1993measureorder}), and two at the mesoscale (modularity\cite{newman2004finding}, in-block nestedness\cite{lewinsohn2006structure,sole2018revealing,palazzi2019antagonistic}). We focus our attention on modular, nested and in-block nested patterns since all of them have been observed prominently in ecology\cite{olesen2007modularity,fortuna2010nestedness,Thebault2010,flores2011statistical} and in information systems\cite{borge2017emergence,sole2018revealing}. 
We quantify the amount of nestedness by means of a global nestedness fitness $\altmathcal{N}$, introduced by Sol\'e-Ribalta {\it et al.}\cite{sole2018revealing}, an overlap measure\cite{almeida2008consistent} that includes a suitable null model. We follow this work as well for the definition and optimisation strategy of in-block nestedness $\altmathcal{I}$. With respect to community analysis, we apply a variant of the extremal optimisation algorithm\cite{duch2005community}, adapted\cite{palazzi2019antagonistic} to maximise Barber's bipartite modularity\cite{barber2007modularity}.


\paragraph{Niche model and population dynamics.}
To perform the numerical simulations, we employ a model that follows a Lotka-Volterra dynamics, with Holling-Type II mutualistic functional response\cite{bastolla2009architecture,suweis2013emergence}:
\vspace{.1cm}
\begin{equation}
\label{eq:dyn}
{\begin{aligned}{\frac {dn_i^U}{dt}}&=n_i^U \left(\rho_i^U - \sum_j \beta_{ij}^Un_j^U + \frac{\sum_k \gamma_{ik}^{UH}n_k^H }{1+ h \sum_k \theta_{ik}^{UH}n_k^H} \right), \\[6pt]
{\frac {dn_i^H}{dt}}&=n_i^H \left(\rho_i^H - \sum_j \beta_{ij}^Hn_j^H + \frac{\sum_k \gamma_{ik}^{HU}n_k^U }{1+ h \sum_k \theta_{ik}^{HU}n_k^U} \right).
\end{aligned}}
\end{equation}
Here, the coupling matrices $\mathbf{\beta}$ and $\mathbf{\gamma}$ define the competitive (within guild) and mutualistic (across guild) interactions, respectively. Both interaction matrices depend linearly on the niche overlap between pairs of species. Additionally, these matrices have a global factor, $\Omega_c$ or $\Omega_m$, which tune the strength of competitive or mutualistic interaction, respectively. The niche profile for each species is modelled as a Gaussian function, and we assume that each user and hashtag is involved in niche relations according to assigned topics of their interest. In particular, a number of $T$ equidistant topics are created on the niche axis. Finally, $\mathbf{\theta}$ is the adjacency matrix. , and $h$ is the handling time of the Holling-Type II mutualistic functional response. See the Supplementary Material for details.

In our information ecosystem, these equations represent a phenomenological way to describe the evolution of the nodes visibility as a function of their interaction. In particular, $n_i^U$ may represent the number \vc{of} instances in which user $i$ is present in other users' screens, while $n_j^H$ may quantify the popularity of a given hashtag $j$. Assuming that preferential attachment mechanisms of various type affect the nodes visibility, $\rho_i^U$ and $\rho_i^H$ model the associated exponential growth (if they are positive). The handling time $h$ effectively models the constraint that users cannot interact with a very large number of hashtag\vc{s} due both to time and character constraints. Due to these limitations, the benefit obtained through mutualistic interactions does not grow monotonically with the number of partners.

Simulations are performed by integrating the system of ordinary differential equations using a fourth-order Runge-Kutta method. Then, we start a rewiring process following the approaches in Suweis {\it et al.}\cite{suweis2013emergence} and Cai {\it et al.}\cite{cai2018dynamic}: at constant time intervals, species will rewire recurrently in order to maximise their individual abundances. We assign the same initial abundance $n_0=0.2$ and intrinsic growth rates  $\rho_U = \rho_H = 1$ to all users and hashtags. Species are considered to suffer extinction when their abundance density is lower than $10^{-4}$. Finally, the handling time $h$, of the Holling-Type II mutualistic functional response is set to $0.1$.

\bibliography{nest}

\section*{Acknowledgments}
M.J.P, A.S-R. and J.B-H. acknowledge the support of the Spanish MICINN project PGC2018-096999-A-I00. M.J.P. acknowledges as well the support of a doctoral grant from the Universitat Oberta de Catalunya (UOC). S.S. thanks the support of UNIPD through ReACT Stars 2018 grant. S.M. and V.C. acknowledge partial financial support from the Agencia Estatal de Investigacion (AEI, Spain) and Fondo Europeo de Desarrollo Regional under Project PACSS Project No. RTI2018-093732-B-C22  (MCIU, AEI/FEDER,UE) and through the Mar\'ia de Maeztu Program for units of Excellence in R\&D (MDM-2017-0711). All authors acknowledge the support to the TEAMS project of the Cariparo Visiting Program 2018 (Padova, Italy). We also thank Joan T. Matamalas from the Universitat Rovira i Virgili for the help with the acquisition of the Spanish elections dataset.

\section*{Author contributions statement}

All authors designed research. A.S-R. and J.M. collected and curated the data. M.J.P., A.S-R. and J.B-H. and performed research. All authors analysed the results. J.B-H., S.S. and S.M. wrote the paper. All authors approved the final version.

\noindent\textbf{Competing Interests} The authors declare no competing interests.

\noindent\textbf{Correspondence} Correspondence and requests for materials should be addressed to J.B-H. \\ (email: jborgeh@uoc.edu).

\end{document}



\title{Supplementary Material\\ Resilience and elasticity of co-evolving information ecosystems}

\author{ Mar\'ia J. Palazzi$^1$, Albert Sol\'e-Ribalta$^{1,2}$, Sandro Meloni$^3$, Violeta Calleja-Solanas$^3$, Carlos A. Plata$^4$,\\ Samir Suweis$^4$ and Javier Borge-Hotlhoefer$^1$ \\
\small{$^1 $Internet Interdisciplinary Institute (IN3), Universitat Oberta de Catalunya, Barcelona, Catalonia, Spain}\\
\small{$^2 $URPP Social Networks, University of Zurich, Zurich, Switzerland}\\
\small{$^3$IFISC, Institute for Cross-Disciplinary Physics and Complex Systems (CSIC-UIB), 07122, Palma de Mallorca, Spain}\\
\small{$^4$Dipartimento di Fisica e Astronomia G. Galilei, Università di Padova, Via Marzolo 8, Padova, 35131, Italy}}

\maketitle

\section{Empirical analysis}
\subsection{Datasets}

The empirical data employed in this work was collected from the online platform \url{www.twitter.com}. Following the analogy with interactions in ecological systems, two types of species are considered: users and hashtags (memes). For each tweet on the different datasets, we only extracted the user's name, the hashtags in the tweets, and the time at which they were posted.

We considered six  events of different nature: the Spanish general elections april 2019 (28A), the 2015 Nepal earthquakes, the 2012 UEFA European Football Championship, the 2014 Catalan self-determination referendum, the 2015 Charlie Hebdo Shooting and the 2014 Hong Kong streets protests. All the datasets excepting the ones from the Spanish general elections and the Catalan self-determination referendum were collected by Zubiaga A. in \cite{zubiaga2018longitudinal}. In the following, we report details concerning these events and the associated Twitter data sets.

\paragraph{Spanish general Elections (April 2019):}
The April 2019 Spanish general elections were held on Sunday, 28 April 2019, to elect the 13th bicameral legislative chambers of the Kingdom of Spain, the $350$ seats in the Congress of Deputies and $208$ out of $266$ seats in the Senate. The observation period started at the beginning of the electoral campaign, on the 12th of April and lasted until the 6th of May, a few days before the beginning of the electoral campaign for the election of the 54  Spanish members of the European Parliament. Hence, the observation period was marked by intense political activity. For this event, we collected a dataset composed of  $3,0107,629$ unique tweets containing at least one hashtag, with a total $124,062$ unique hashtags and $1,883,468$ users.  The dataset was collected by selecting all the tweets containing at least one of a total set composed of 300 relevant keywords that could be either user names or hashtags related to the electoral process, i.e, names of candidates, electoral actvities (debates, meetings) and name of the parties involved, etc.  The collection of this dataset was carried out in collaboration with Joan Tomàs Matamalas. 

\paragraph{Nepal Earthquake (April-May 2015):}
The next dataset taken into consideration for analysis corresponds to an unexpected event, specifically, a series of earthquakes registered in Nepal in 2015.  The first earthquake occurred on the 25 of April 2015, registering around 9000 casualties. This event was followed by several continued aftershocks, with a major aftershock of similar magnitude of the first quake, registered on May 12th. Given the unpredictable nature of this type of event, we have focused on the study of the second major earthquake. The observation period covers a total of six days, from 8 to 14 of May, a few days after the second aftershock.  The dataset contains $1,918,045$ unique tweets containing at least one hashtag, with a total $35,795$ unique hashtags and $810,744$ users.  The dataset was collected by selecting all the tweets containing at least one of the following hashtags or keywords: \textit{nepal}, \textit{earthquake}, \#\textit{nepalearthquake}. 

\paragraph{Catalan self-determination referendum (Nov 2014):}
The second event corresponds to the  Citizen's Participation Process on the Political Future of Catalonia, a popular consultation about the process of  independence of Catalonia from the Spanish Kingdom. The consultation was held on Sunday, 9 November 2014, after the approval decree was signed by the president of Catalonia on September 27 of the same year. The dataset contains $220,364$ unique tweets containing at least one hashtag, with a total $18,116$ unique hashtags and $78,270$ users, ranging from September 1st to November 13 of 2014. In a similar manner to the Spanish election dataset, this dataset was collected by selecting all the tweets containing at least one of a preselected set of $\approx 70$ hashtags and $\approx 50$ Twitter accounts related to the referendum process and the Catalan independence movement. 

\paragraph{European football championship (2012):}
In third place, we considered the 2012 UEFA Football Championship, an European championship for men's national football teams. The tournament was held between 2 June and 1 July  of 2012, and co-hosted by Poland and Ukraine. The observation period started a day before of the quarter-finals, on 19 June and lasted until the 4th of July, right after the final game. It contains $3,907,418$ unique tweets containing at least one hashtag, with a total $ 147,646$ unique hashtags and $1,325,631$ users. This dataset was collected by selecting all the tweets containing the hashtag \#\textit{euro2012}.

\paragraph{Hong Kong protests (Sept-Oct 2014):}
An additional dataset taken into consideration corresponds to a series of streets protests that took place in Hong Kong from September to December 2014. The protests, are often referred to as the Umbrella Movement or Occupy movement. The protests were initiated after a proposal from the  Standing Committee of the National People's Congress to reform the electoral law. The dataset contains $826,194$ unique tweets containing at least one hashtag, with a total $30,105$ unique hashtags and $239,432$ users. The observation period started on 27th of September, right after the  protests escalated, resulting in several people detained, until October 10. The dataset was collected by selecting all the tweets containing at least one of the following hashtags or keywords: \#\textit{hongkong}, \#\textit{umbrellamovement}, \#\textit{occupycentral}, \#\textit{hongkongprotests},  \#\textit{occupyhongkong}. 

\paragraph{Charlie Hebdo Shooting (Jan 2015):}
The last dataset taken into consideration for analysis also corresponds to an unexpected event, specifically, the shooting perpetrated at the offices of the french magazine Charlie Hebdo, on January 7 2015.  On the morning of January 7 of 2015, two heavily armed brothers forced their entry into the magazine offices, killing 12 people and injuring 11 more. The dataset contains $6,002,087$ unique tweets containing at least one hashtag, with a total  $102,799 $ unique hashtags and  $2,001,826$ users. The observation period started on the 8th of January, right after the shooting took place and lasted until the 10th of January, after the two main suspects were killed. The dataset was collected by selecting all the tweets containing at least one of the following hashtags or keywords: \#\textit{jesuischarlie},  \#\textit{charliehebdo}, \textit{charlie hebdo} \textit{paris}. 




\begin{table}[ht]
\caption{Summary of our datasets.}
\begin{center}
\begin{tabular}{c|c|c|c|c|c}
Event & Data length & Total days & Tweets & Users & Hashtags  \\\hline
2019 Spanish general elections &April 12 - May 6&24& 30,107,629 &1,883,468  &124,062\\  \hline
2015 Nepal Earthquake &May 8-14&6& 1,918,045   & 810,744 &35,795  \\\hline 
2014 Catalan referendum &Sep 2 - Nov 12& 10& 220,364 & 78,270 &18,116  \\  \hline
2012 UEFA football championship &Jun 19 - July 4&15 & 3,907,418 &  1,325,631 &  147,646\\   \hline
2014 Hong Kong Protests &Sep 27 - Oct 7&10&  826,194  &  239,432 & 30,105  \\\hline
2015 Charlie Hebdo shooting  &Jan 8-9 &2& 6,002,087 & 2,001,826&102,799   \\\hline
\end{tabular}
\end{center}
\label{table1}
\end{table}%

\subsection{Matrix construction}
As it is explained in the main text, we attempt to account for how the interactions between users and hashtags change over time.
Prior the construction of the interaction matrices, we performed a  selection criteria that allowed us to capture the structural changes of the data in a smooth way, and, at the same time, reducing the computational cost. 

\begin{figure}[h!]
\centering
\topinset{\bfseries}{\includegraphics[width=.85\textwidth]{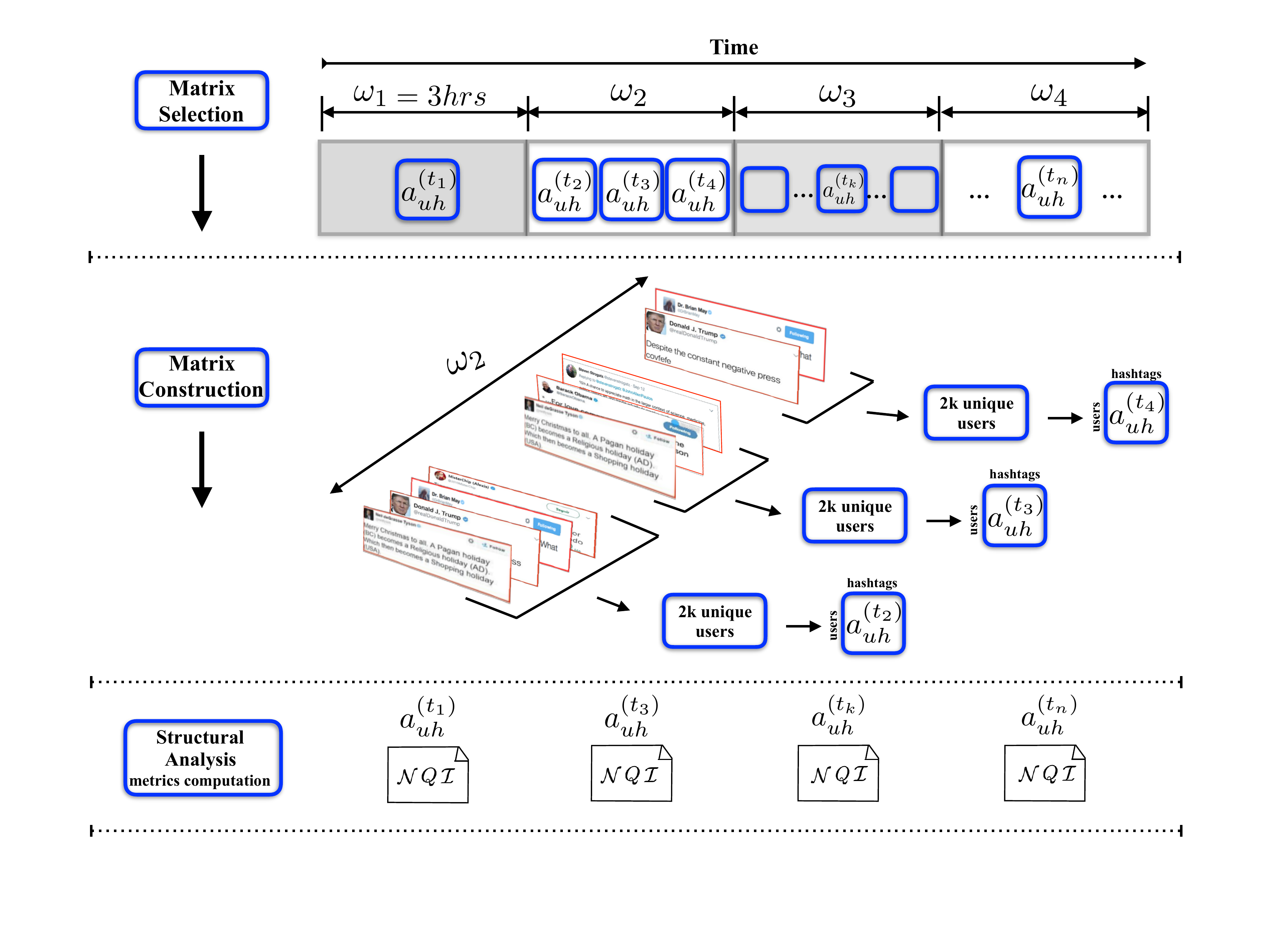}}{0.07in}{.2in}
\vspace{-.6cm}
\caption{\textbf{Schematic representation of the implemented methodology in the analysis of empirical data.}The applied methodology comprises three steps: Selection and construction of the adjacency matrices (top and middle rows) , structural analysis of the selected matrices by means of nestedness, modularity and in-block nestedness (bottom row).}\label{fig:fig1}
\end{figure}

For each dataset, we split the timestream into chunks according to non-overlapping time windows with three hours of duration $\omega=3$h, Fig.~\ref{fig:fig1} top row. For each chunk, we built matrices $a^{(t)}_{uh}$ containing the $2000$ most active unique users and a variable number of hashtags, depending on the amount produced by those 2000 users \cite{borge2017emergence}. Each cell in the matrices $a^{(t)}_{uh}$ is equal to $1$ if user $u$ has posted a message containing the hashtag $h$ at least once, and $0$ otherwise. Note that each matrix will have a different duration, spanning from a few minutes during the times of high activity (when an event is taking place), to the total duration of the time window.  For each one of these 3-hour chunks, we select the matrices that are closer to the middle of the time window, e.g., $a^{(t \approx \omega/2)}$ to perform the structural analysis. 

It is also important to highlight that the $a^{(t)}_{uh}$ matrices may not contain the same nodes across $t$: as time advances, users join (disappear) as they start (cease) to show activity; the same applies for hashtags, which might or might not be in the focus of attention of users. This volatile situation is quite normal in time-resolved ecology field studies \cite{alarcon2008year,petanidou2008long,diaz2010changes}, where the accent is placed on the system's dynamics --rather than individual species.

Around the periods of high activity --on the onset of the events-- the procedure is repeated considering time windows of 15 minutes of duration.


\subsection{Structural analysis}
%
In this section, we introduce the two structural arrangements mentioned throughout the paper.

\begin{figure}[h]	
\centering
                \includegraphics[width=0.8\textwidth]{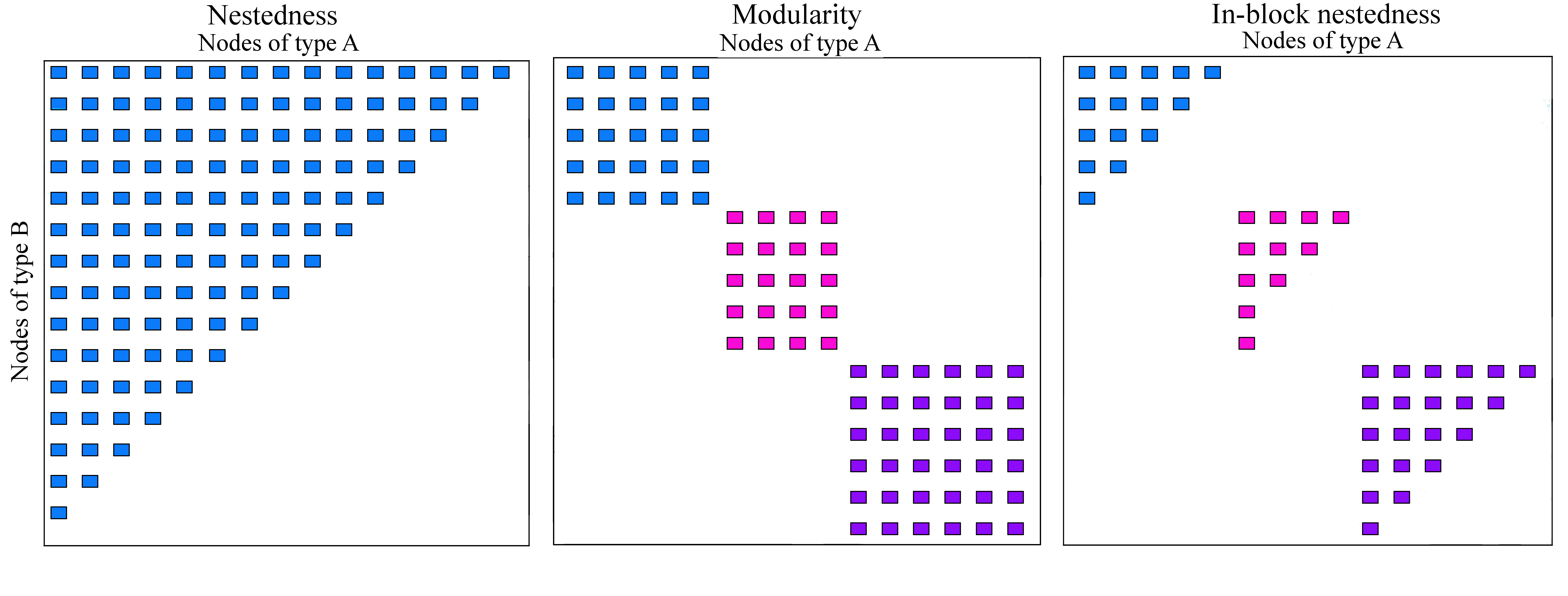} 
	\caption{\textbf{Idealised examples of the structural organisation studied in the paper.} Left panel shows a perfectly nested organisation. Middle panel the adjacency matrix of modular network and right Panel an idealised network with in-block nested structure.}
	\label{fig_examplePaterns}        
\end{figure}

\subsubsection{Nestedness}

The concept of nestedness appeared, in the context of complex networks, over a decade ago in systems ecology \cite{bascompte2003nested}, and was previously introduced as a way to describe the patterns of distribution of species in isolated habitats \cite{patterson1986patterson}. In structural terms, a perfect nested pattern is observed when specialists (nodes with low connectivity) interact with proper nested subsets of those species interacting with generalists (nodes with high connectivity), see Figure \ref{fig_examplePaterns} (left). 

Since the seminal work of Bascompte {\it{et al}}, the nested pattern has been sought (and found) in many different contexts \cite{bustos2012dynamics,Saavedra2011,borge2017emergence}. At the same time, it has been implicated as a key pattern in the dynamical proccess of ecological systems \cite{bastolla2009architecture,Thebault2010,Suweis2013emergence}.  For this reason, scholars have dedicated many efforts to quantify nested patterns in real systems. In first place, we have measures based on counting misplaced relations to complete a perfect upper triangular nested structure on the adjacency matrix $A$, such as the Nested Temperature (NT) measure, introduced by Atmar and Patterson \cite{atmar1993measureorder}. Second, we have overlap metrics like the Node Overlap and Decreasing Fill (NODF), developed by Almeida-Neto {\it et al.} \cite{almeida2008consistent}, which considers the amount of common neighbours between every two pair of nodes in the adjacency matrix $A$.  

Here, we quantify the amount of nestedness in information networks by employing an overlap metric introduced by Sol\'e-Ribalta {\it et al.}\cite{sole2018revealing}:
\begin{center}
\begin{equation} 
\altmathcal{{N}} = \frac{2}{N_r+N_c} \left \{  \sum_{i,j}^{N_r} \left[  \frac{\mathit{O}_{ij}- \langle \mathit{O}_{ij}\rangle}{k_{j}(N_r - 1)} \Theta( k_{i} -k_{j} )\right]  + \sum_{l,m}^{N_c} \left[  \frac{\mathit{O}_{lm}- \langle \mathit{O}_{lm}\rangle}{k_{m}(N_c - 1)} \Theta( k_{l} -k_{m} )\right]  \right \},
 \label{eq_nest}
 \end{equation}
  \end{center}
where $N_r$ and $N_c$ correspond to the number of rows and column nodes, respectively. The values$O_{ij}$ (or  $O_{lm}$) measure the degree of links overlap between rows (or columns) node pairs; $k_i$,$k_j$ corresponds to the degree of the nodes $i$,$j$,  and $\Theta(\cdot)$ is a Heaviside step function that guarantees that we only compute the overlap between pair of nodes when $k_i \geq  k_j $. Finally, $\langle \mathit{O}_{ij}\rangle$ represents the expected number of links between row nodes $i$ and $j$ in the null model, and is equal to $\langle \mathit{O}_{ij}\rangle=\frac{k_ik_j}{N_r}$. 

\subsubsection{Modularity}

Modular structure is a rather ubiquitous mesoscale architecture \cite{zachary1977information,guimera05,adamic2005political,fortunato2010community} and implies that nodes are organised forming groups, i.e.~devoting many links to nodes in the same group, and fewer links towards nodes outside \cite{radicchi2004defining}, see Fig.~\ref{fig_examplePaterns}(middle). Given the huge number of possible ways to partition a graph into groups  an exhaustive assessment of every partition's fitness is unfeasible. Hence, scholars have developed several algorithms that are able to find fairly good approximations or (sub)optimal partitions, by means of the optimization of a fitness function \cite{newman2004findin, duch2005community,blondel2008fast,Fortunato2016} . Here, we search for a (sub)optimal modular partition of the nodes by applying the extremal optimisation algorithm \cite{duch2005community} to maximise Barber's\cite{barber2007modularity} modularity, which is an extention of the original formulation introduced by Newman \cite{newman2004finding}, to bipartite networks:
\begin{center}
\begin{equation} 
Q = \frac{1}{L} \displaystyle\sum^{N_r}_{i=1}\sum^{N_c}_{j=1} \left(\tilde{A}_{ij}-\tilde{p}_{ij}\right)\delta(\alpha_{i}^{r},\alpha_{j}^{c})
  \end{equation}
  \end{center}
where $L$ is the number of interactions (links) in the network, $\tilde{A}_{ij}$ is the adjacency matrix which denotes the existence of a link between rows and columns nodes $i$ and $j$, $\tilde{p}_{ij}=k_{i}k_{j}/L$ is the probability that a link exists by chance, and $\delta(\alpha_i,\alpha_j)$ is the Kronecker delta function, which takes the value 1 if nodes $i$ and $j$ are in the same community, and 0 otherwise. 
\subsubsection{In-block nestedness}

Nestedness and modularity are emergent properties in many systems, but it is rare to find them in the same system. This apparent incompatibility has been noticed and studied, and it can be explained by different evolutive pressures: certain mechanisms favour the emergence of blocks, while others favour the emergence of nested patterns. Following this logic, if two such mechanisms are concurrent, then hybrid (nested-modular) arrangements may appear. Hence, the third architectural organisation that we consider in our work refers to a mesoscale hybrid pattern, in which the network presents a modular structure, but the interactions within each module are nested, i.e. an in-block nested structure, see Figure~\ref{fig_examplePaterns} (right). This type of hybrid or ``compound" architectures was first described in Lewinsohn {\it et al}.\cite{lewinsohn2006structure} and has been further explored in the last decade \cite{flores2013multi,beckett2013coevolutionary,sole2018revealing}.

Using the formulation developed in \cite{sole2018revealing}, the degree of in-block nestedness of a network $\altmathcal{I}$ can be computed as

\begin{center}
\begin{equation}
\altmathcal{I} = \frac{2}{N_r+N_c} \left \{  \sum_{i,j}^{N_r} \left[  \frac{\mathit{O}_{ij}- \langle \mathit{O}_{ij}\rangle}{k_{j}(C_{i} - 1)} \Theta( k_{i} -k_{j} ) \delta (\alpha _{i}, \alpha _{j}) \right] +  \sum_{l,m}^{N_c} \left[  \frac{\mathit{O}_{lm}- \langle \mathit{O}_{lm}\rangle}{k_{m}(C_{l} - 1)} \Theta( k_{l} -k_{m} ) \delta (\alpha _{l}, \alpha _{m}) \right] \right \},
\label{eq_inb}
\end{equation}
\end{center}
where $C_{i}$ accounts for the number of nodes in the same guilds of $i$ and that belong to the same community.  Worth highlighting this hybrid structure reframes nestedness, originally a macroscale feature, to the mesoscopic level. In this sense, by definition, $\altmathcal{I}$ reduces to $\altmathcal{N}$ when the number of blocks is 1.

\section{Dynamical model}

\subsection{Niche model}
The model is developed for a bipartite network that contains interacting ``species" in two classes or guilds (denoted $U$ and $H$, in analogy with users and hashtags). For each species $i$ we assign a niche profile, which is formulated as a Gaussian function $G_i(s)$ with width $\sigma_i$ -- for simplicity, we assign the same niche width to all the species-- and its center position $\overline{s}_i$ are chosen selected within the interval $[0, 1]$ on a niche axis, with fixed boundary conditions.
In the original formulation, the center positions of the niche profiles were randomly distributed along the niche axis. In this work, we  assume that each user and hashtag is involved in niche relations according to specified topics of its interest. In particular, a number of $T$ topics are created equidistant on the niche axis. The niche center of the $N_{U}$ users and $N_{H}$ hashtags  are set in the vicinity of each topic. 

\subsection{Mutualistic and competitive interactions.} 
The species are involved in cross-guild mutualistic interactions and into competitive interactions with all the nodes in its own guild proportional to the niche overlap. We define the niche overlap $G_{ij}$ of a pair of nodes $i$ and $j$  as: 

\begin{equation}
G^{gg'}_{ij}= \int G^{g}_i(s)G^{g'}_j(s)ds 
\end{equation}
with $g$ and $g'$ denoting the guild of the considered species, either users or hashtags.

\subsubsection{Mutualism}
Following this, we define the mutualistic interaction matrix as:
\begin{equation}
\label{eq:smut}
\text{Mutualism: } \gamma_{ik}^{UH} = \Omega_m \cdot \theta_{ik} \cdot G_{ik}^{UH}
\end{equation}
where $\theta_{ik}$ is the adjacency matrix, with entries equal to $1$ if $i$ and $k$ interact, and $0$ otherwise, and $\Omega_m$  is the intensity coefficients for the mutualistic interaction. 
\subsubsection{Competition}
Regarding the competitive interactions, in the context of information ecosystems we distinguish two levels of competition. At the local level, users (hashtags) in the same topic compete to gain visibility among those with related interests (meaning). At the aggregate level, a given topic strives to prevail among other topics. In order to capture this double competition as a trade off between both tendencies in our model, we need to redefine the competitive interaction matrix as:
\begin{equation}
\label{eq:scom}
\text{Competition: } \beta_{ij}^{U} =
\begin{cases}
1  &  \text{if $i=j$} \\
\Omega_c \left[ \lambda(1- G_{ij}^{UU}) + (1-\lambda)G_{ij}^{UU} \right], & \text{otherwise},
\end{cases}
\end{equation}
where $\Omega_c$  is the intensity coefficients for the competitive interaction and $\lambda \in[0,1]$ is the inter-intra topic competition parameter, the same definition applies to the competitive interactions among hashtags. For the case $\lambda=1$, the competitive matrix neglects the competition among users belonging to the same topic. The case when $\lambda=0$ corresponds to the original formulation \cite{cai2018dynamic}.

\subsection{Population dynamics}
The species abundances evolve according to a set of Lotka-Volterra equations with Holling-Type II mutualistic functional response with handling time $h$:
\vspace{.1cm}
\begin{equation}
\label{eq:sdyn}
{\begin{aligned}{\frac {dn_i^U}{dt}}&=n_i^U \left(\rho_i^U - \sum_j \beta_{ij}^Un_j^U + \frac{\sum_k \gamma_{ik}^{UH}n_k^H }{1+ h \sum_k \theta_{ik}^{UH}n_k^H} \right)\\[6pt]
{\frac {dn_i^H}{dt}}&=n_i^H \left(\rho_i^H - \sum_j \beta_{ij}^Hn_j^H + \frac{\sum_k \gamma_{ik}^{HU}n_k^U }{1+ h \sum_k \theta_{ik}^{HU}n_k^U} \right).
\end{aligned}}
\end{equation}
where the handling time $h$, of the Holling-Type II mutualistic functional response is set to $0.1$. Simulations were performed by integrating the system of ordinary differential equations using a fourth-order Runge-Kutta method. 

\subsection{Optimization process}
We consider a rewiring adaptation process that follows the approaches in \cite{Suweis2013emergence,cai2018dynamic}. At constant time intervals, species will rewire recurrently in order to maximize their individual abundances.

\subsubsection{Rewiring:} At each time step $t=mT$ ($m$ is a positive integer and $T$ is the integration time), a random species $u$, with a least one link, is selected and rewired to a randomly selected species $h'$, removing one of its previous links $h$, with probability $p_{uh} \propto 1-k_h^{-1}$. 
The rewiring probability is defined in such a way that the larger the species' degree is, the more prone to losing links. Once the rewiring is completed,  we recalculate the mutualistic interaction factor of the new pair of nodes $\gamma_{ij'}=\Omega_m \cdot\theta_{ij'} G_{ij'}$ and integrate the dynamics according to Eq. \ref{eq:sdyn}, until the abundances of all species reach an equilibrium (integration time $T$ is set sufficiently large).

\subsubsection{Link recovery:} At the end of each time step $t$, we compare the actual abundance of species $u$ with its previous value. If the current abundance is greater than the previous value, the current (new) link is kept; otherwise, the previous one is recovered. Note that, in the case of abundance loss, only the connections are rolled back to the situation in $t-1$; however, the vectors of abundances continue from their current state, $\vec{n}^U(t)$ and $\vec{n}^H(t)$.


\subsection{Introduction of external events}
Finally, we wanted to explore how the system responds to the introduction of external events that temporarily shifts the population's attention. We modelled this situation as the change of every user's niche center towards a single common topic for a limited period of time. After that period of time, users slowly were moved back to their original niche centers, i.e. back to their respective topics.  

\begin{figure}[h!]
\centering
\topinset{\bfseries (a)}{\includegraphics[width=.47\textwidth]{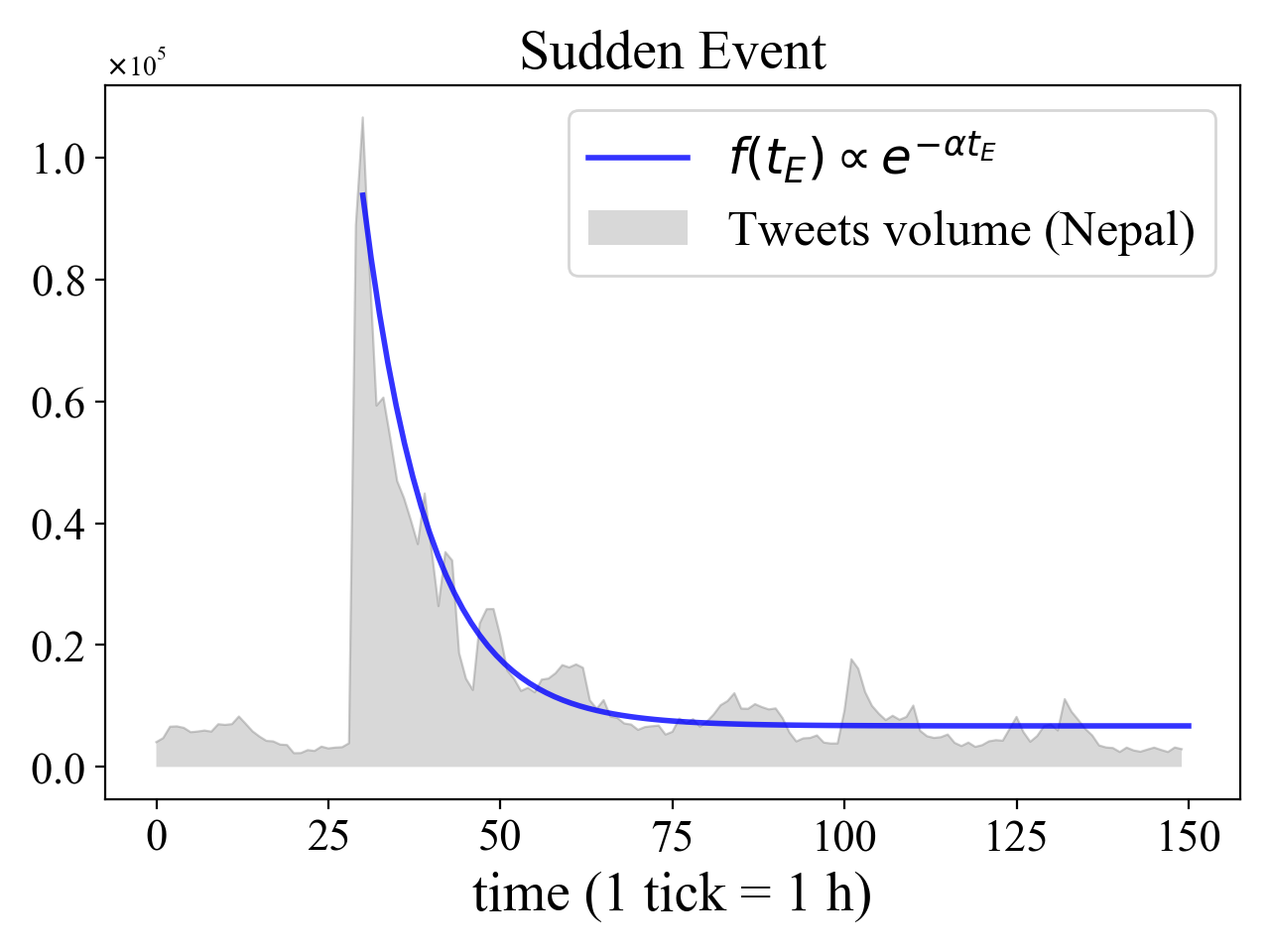}}{0.075in}{-.9in}
\topinset{\bfseries (b)}{\includegraphics[width=.47\textwidth]{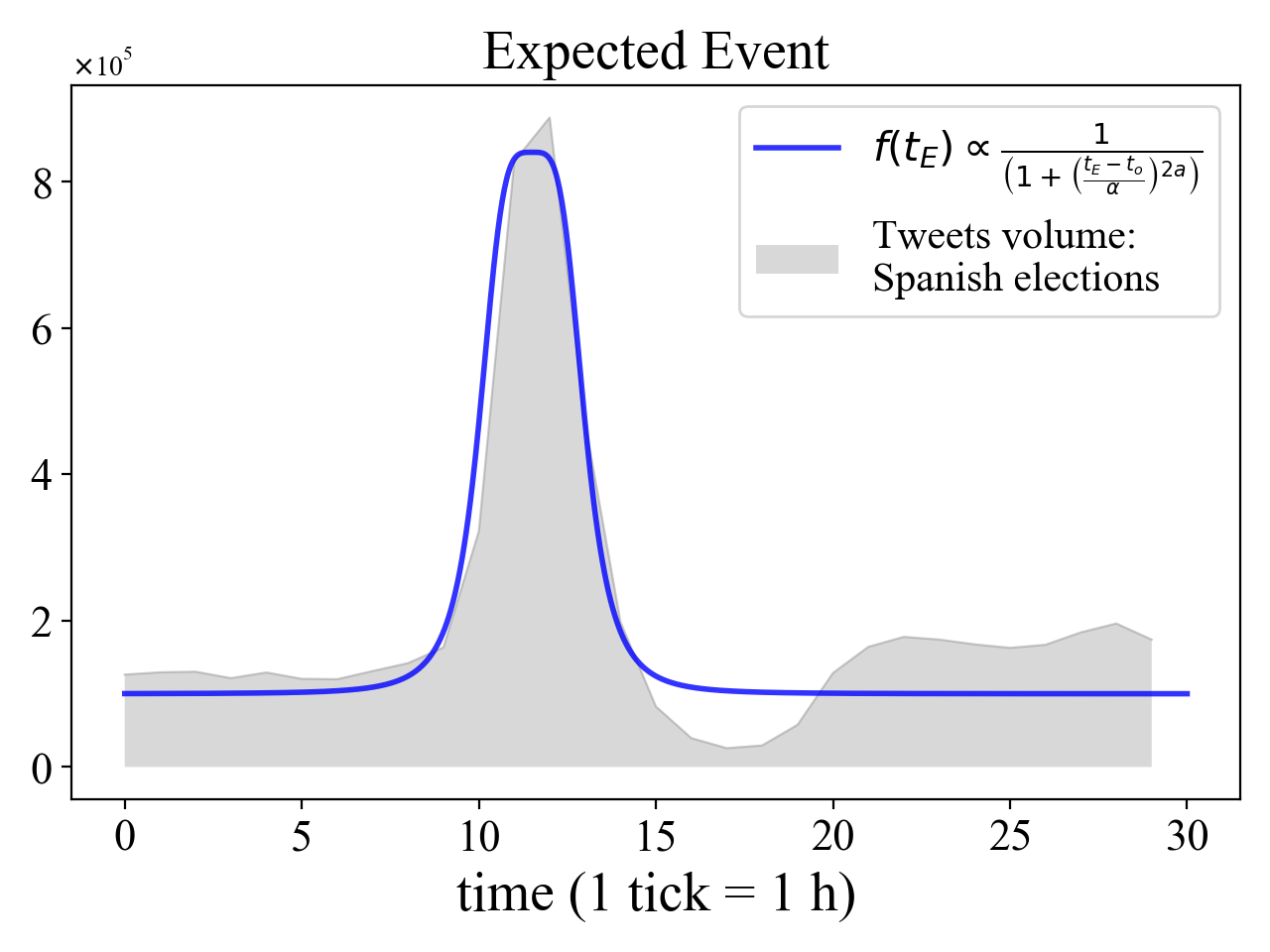}}{0.075in}{-.9in}
	\caption{\textbf{Representation of the two different type of events included in our model}. Panel (a) shows the results a sudden and unexpected event, while panel (b) correspond to an expected event. To ease comparison with real scenarios, in both cases, $f(t_E)$ was shifted in order to align with the maximum and baseline activity of two empirical datasets (shadowed gray areas).}
	\label{fig_exampleevents}        
\end{figure}

An event modifies each users' niche in the following way:
\begin{equation}
G_{i}^{E}(s)=  [1-f(t_{E}) ]G_i(s)   + f(t_{E})G^{E'}(s),
\end{equation}
That is, a user's niche is now the composition of two Gaussian niches: one corresponding to the general event $E$ (defined as a new niche profile $G^{E'}(s)$ centered at $\overline{s}_E$ and width $\sigma_{E}$), and the original one corresponding to the user's intrinsic interests $G_i(s)$. In this formulation, $f(t_{E})$ is the function that governs the growth and decay of the external event, depending on the time $t_{E}$ since its onset. In this work, we modelled two profiles, see Fig. \ref{fig_exampleevents}, along the lines of Lehmann {\it et al.} \cite{lehmann2012dynamical}.

\subsubsection{{Sudden event}}\label{sud_ev}
The first event considered for study was modelled as a sudden and unexpected one. In this case $f(t_{E})$ take the form
\begin{equation}
   f(t_{E}) = e^{-\alpha t_{E}}
\end{equation}
where $\alpha$ is the decay constant. Note that, at the onset of the event ($t_{E} = 0$), all users are focused on the same topic, and their niche overlap will be maximum. For sufficiently large $t_{E}$, namely $t_E \gg \alpha^{-1}$, the influence of the event becomes negligible. 

\subsubsection{Expected event}
In second place, we considered an expected event. In this case, the attention of users will slowly moves towards the one of the event, that is expected to happen an a specific time, in which the user's attention will be maximal. Here, $f(t_{E})$ has the form
\begin{equation}
f(t_{E}) = \frac{1}{\left(1 + \left(\frac{t_E - t_o}{\alpha}\right)^{2a}\right)},
\end{equation}
where $a$ and $\alpha$ are the parameters that regulate the width of the function and the duration of the plateu, while $t_o$ specifies the location of the function peak, note that at $t_E=t_o$ the users niche overlap will be maximum. Again, we modelled the event such that for sufficiently large $t_{E}$, the influence of the event becomes negligible. 


\section{Supplementary Results}
\subsection{Empirical results}
In this section we present four additional results corresponding to different portions of Twitter activity. For the sake of consistency, once again, we analyse these datasets  monitoring the system's modularity \cite{newman2004finding} ($Q$), nestedness \cite{patterson1986patterson,atmar1993measureorder,bascompte2003nested} ($\altmathcal{N}$) and in-block nestedness ($\altmathcal{I}$) \cite{sole2018revealing}. 

\begin{figure}[h!]
\centering
\topinset{\bfseries (a)}{\includegraphics[width=.47\textwidth]{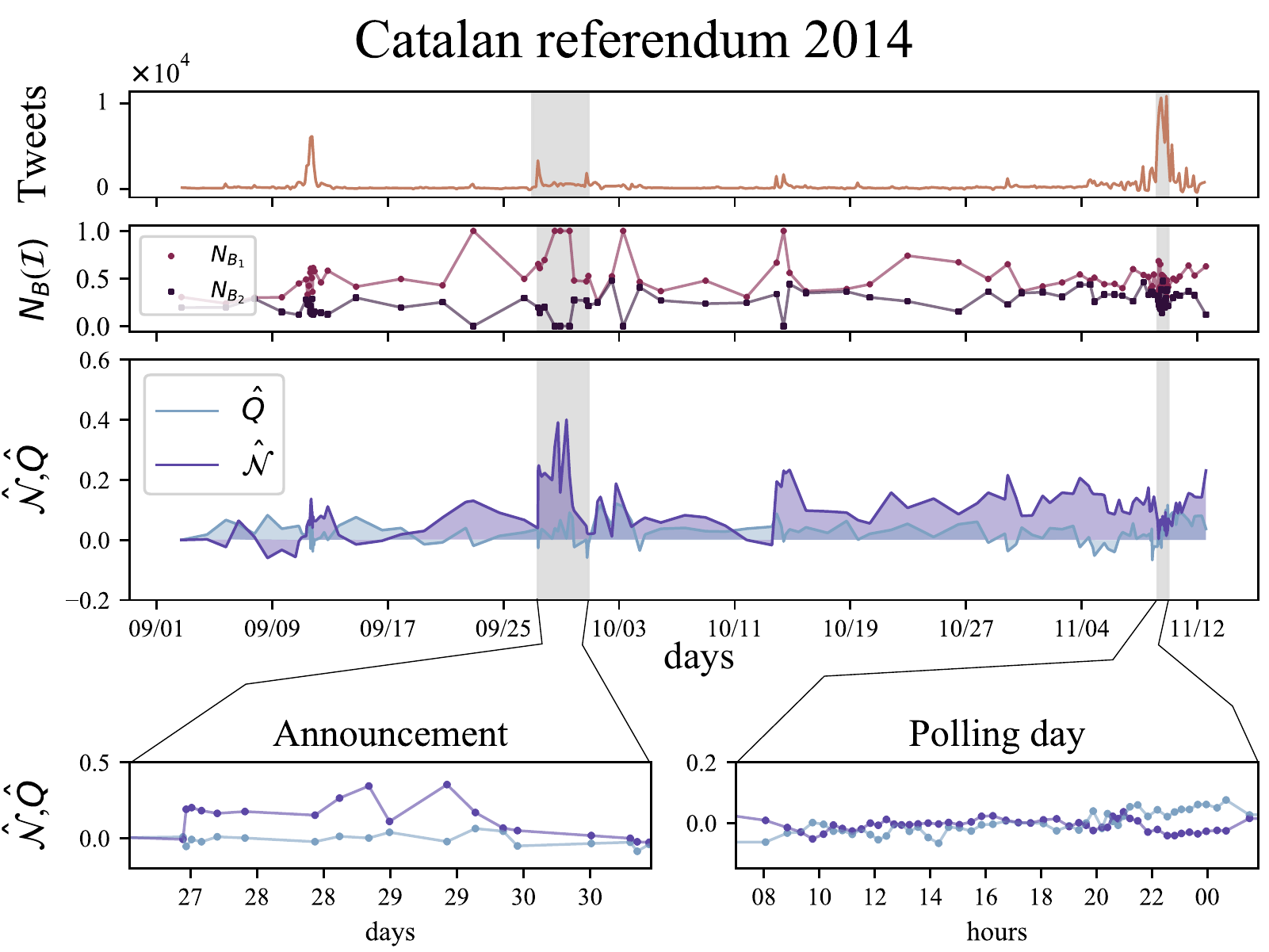}}{0.075in}{1.4in}
\topinset{\bfseries (b)}{\includegraphics[width=.47\textwidth]{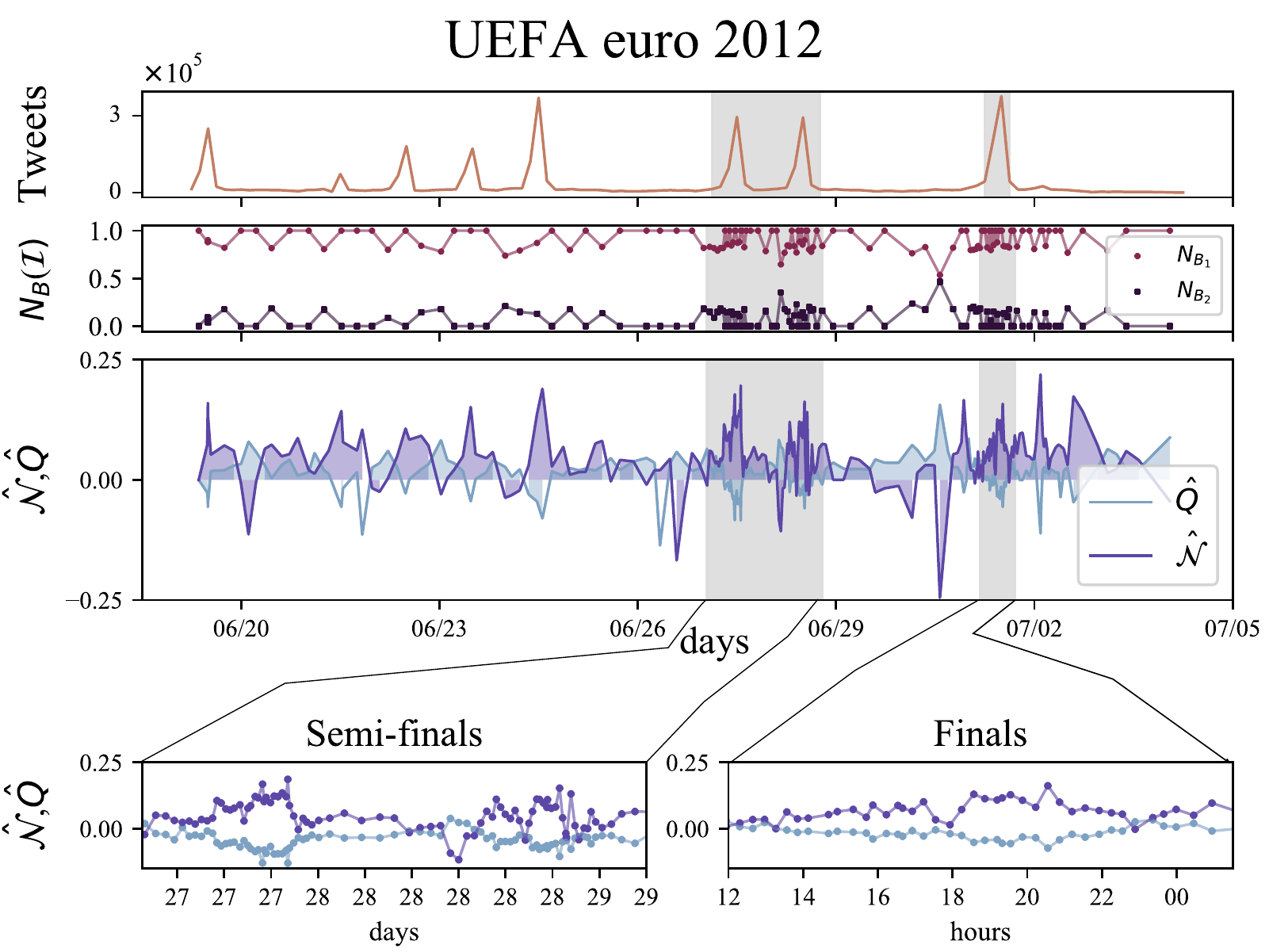}}{0.075in}{1.4in}\\
\topinset{\bfseries (c)}{\includegraphics[width=.47\textwidth]{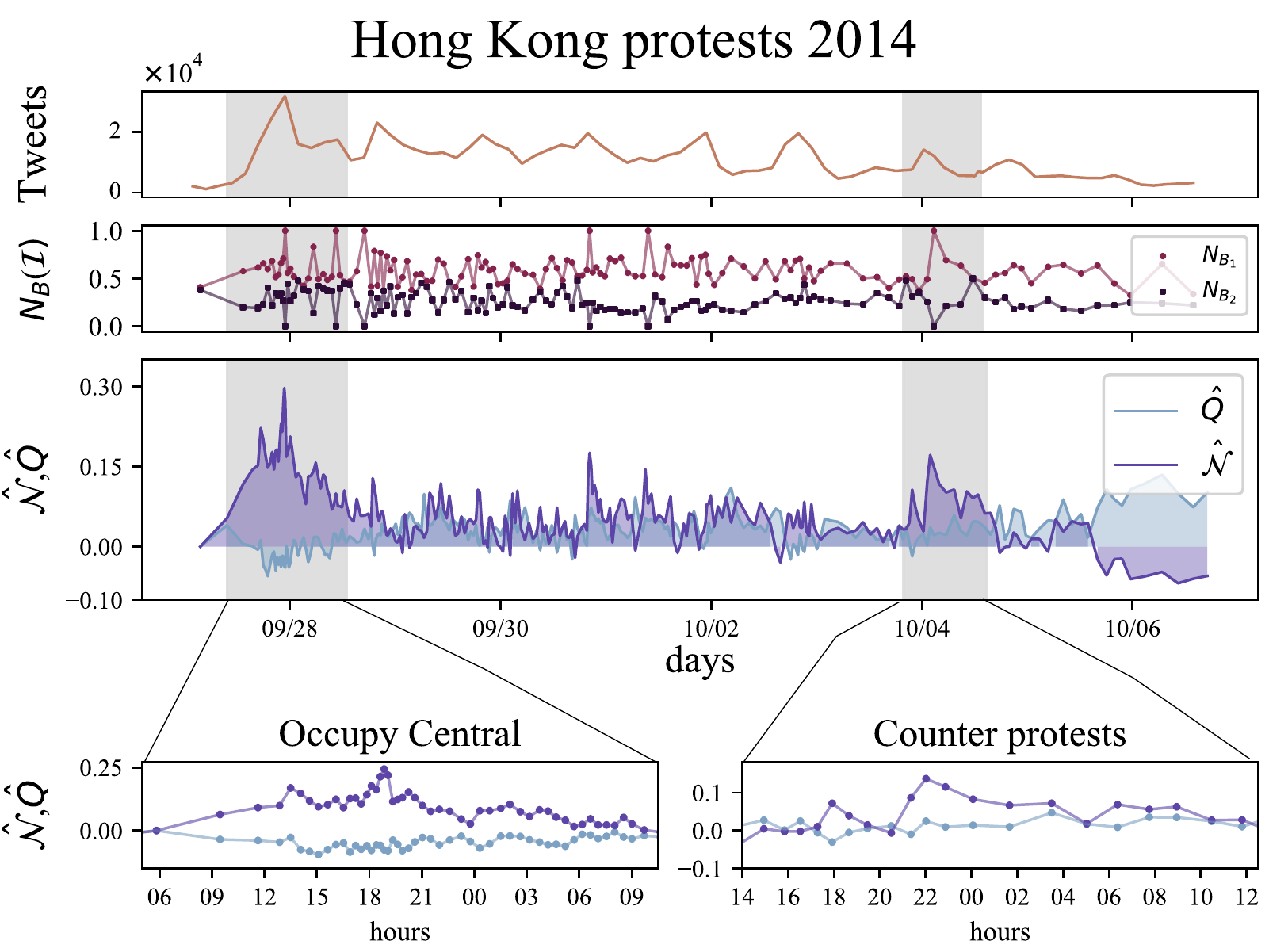}}{0.075in}{1.4in}
\topinset{\bfseries (d)}{\includegraphics[width=.47\textwidth]{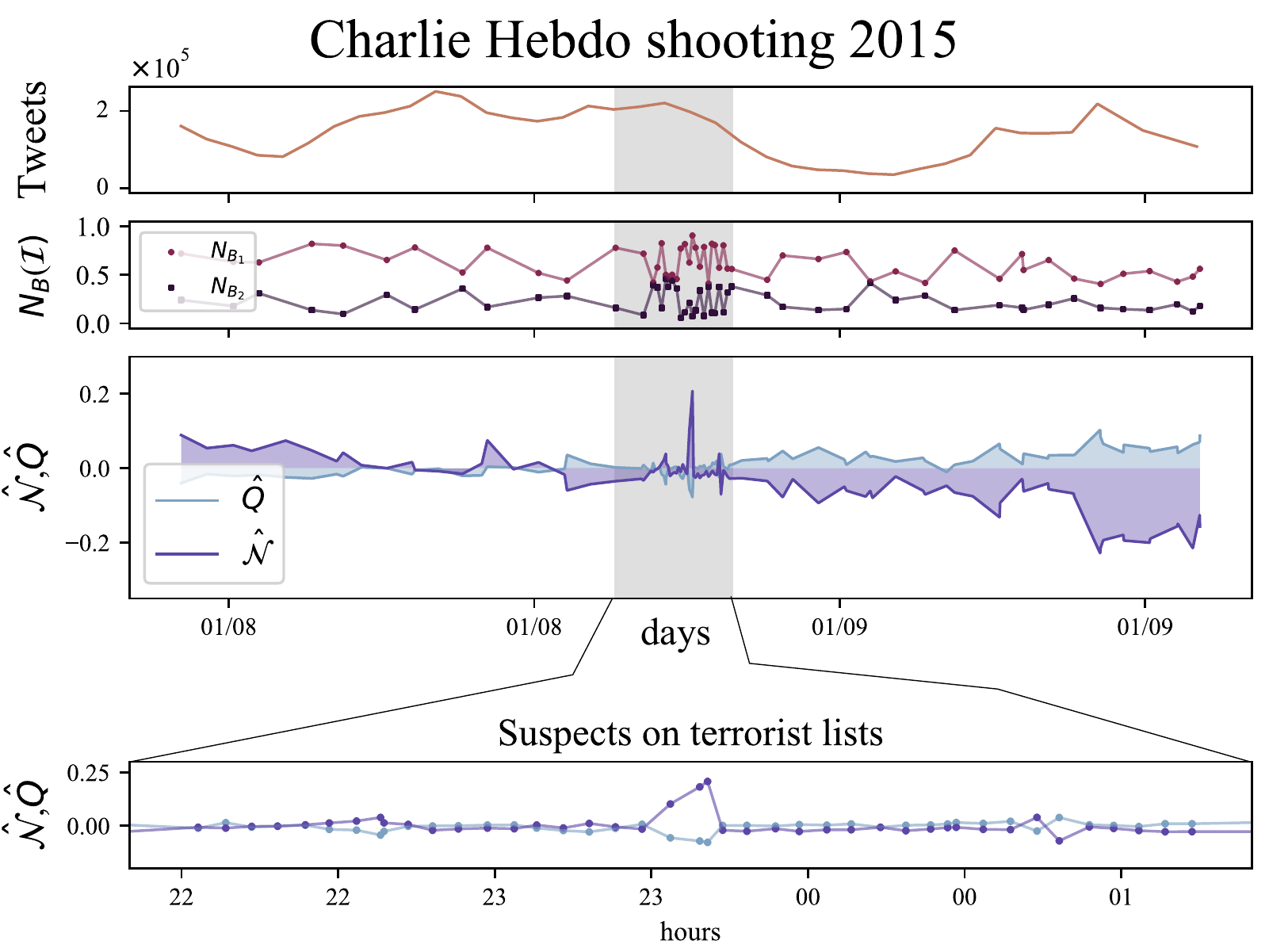}}{0.075in}{1.4in}
\caption{{\bf Structural measures over time for four different datasets.}  Panel (a) corresponds to  the Catalan self-determination referendum held in 2014, panel (b) corresponds to  the 2012 UEFA Football Championship, panel (c) corresponds to a series of streets protests ocurred in Hong Kong in 2014. Finally, panel (d) correspond to the Charlie Hebbdo shooting on 2015. In accordance with empirical results presented in the main text, here we observe how a block organisation dominates the system, reflecting the separate interests of users, until external events induce large-scale attention shifts, which rearrange completely the observed architecture towards a macroscale nested pattern. Once again, we highlight specific time windows in each dataset with some identifiable event happening in them.}
\label{fig:empirical}
\end{figure}

Figure~\ref{fig:empirical} shows the evolution of $Q$ and $\altmathcal{N}$ for the Catalan self-determination referendum from 2014, the 2012  European football championship, the 2014 Hong Kong street protests and the 2015 Charlie Hebdo Magazine shooting. The duration of the snapshot was adjusted to provide a better visualization of the structural transitions during the different events, and highlighted the location of the events in the main panels of each plot. Some of these events are pointed out in the pair of insets in each figure.

Overall, we observe that for all different datasets the behaviour is in qualitative agreement with the ones presented at the main text. First, the anticorrelated behaviour between global nestedness and modularity is preserved. Further, in each case, regardless the nature of the different datasets, we observe a smooth transition into self-similar nested arrangements, which develop in accordance to the level of fragmentation of the surrounding conditions, i.e this transition is linked to external events (second row in all panels). The different datasets, regardless of their nature, lie along the lines of the different classes of collective attention described in Lehmann {\it et al.} \cite{lehmann2012dynamical}. The highly fluctuating pattern in Fig.~\ref{fig:empirical}(b), corresponding to the UEFA championship, is  due to the periodicity in which football games happen throughout the competition, with a slowdown by the end of the period when only the semifinal and final game are left.

Although mentioned in Section~1 of this document, it worth highlighting that different data acquisition procedures employed to build the analysed datasets. The Spanish elections and Catalan referendum datasets were collected from a rich collection of hashtags and keywords that were manually chosen following the evolution of the event, even introducing new hashtags --or keywords-- as the event unfolded. In contrast, the rest of the datasets were collected from a small set of hashtags (often just one) \cite{zubiaga2018longitudinal}, resulting in the presence of ``super"-generalist memes during all the stages of the discussion. Regardless of the possible biases induced by the presence of these ``super"-generalist memes in some of the datasets, many (possibly most) important hashtags emerging at later stages are captured as well, since they tend to co-occur with the original chosen keyword. Thus, we were able to capture the different states of collective attention, from fragmented to global stages of public consensus.

\subsection{Numerical results}

To generalize the results presented in the main text, and to explore if the observed structural transitions are sensitive to changes in the model's local parameters, we perform controlled numerical experiments at the stable stationary state on the $(\Omega_m, \Omega_c)$ parameter space, for different values of the inter-intra competition parameter $\lambda$. We set both $\Omega_m$  and $\Omega_c$ in the interval $[0.1, 0.4]$ and perform simulations for 1200 different combinations of these parameters, for each value of $\lambda$. 

\begin{figure}[h!]	
\centering
                \includegraphics[width=0.75\textwidth]{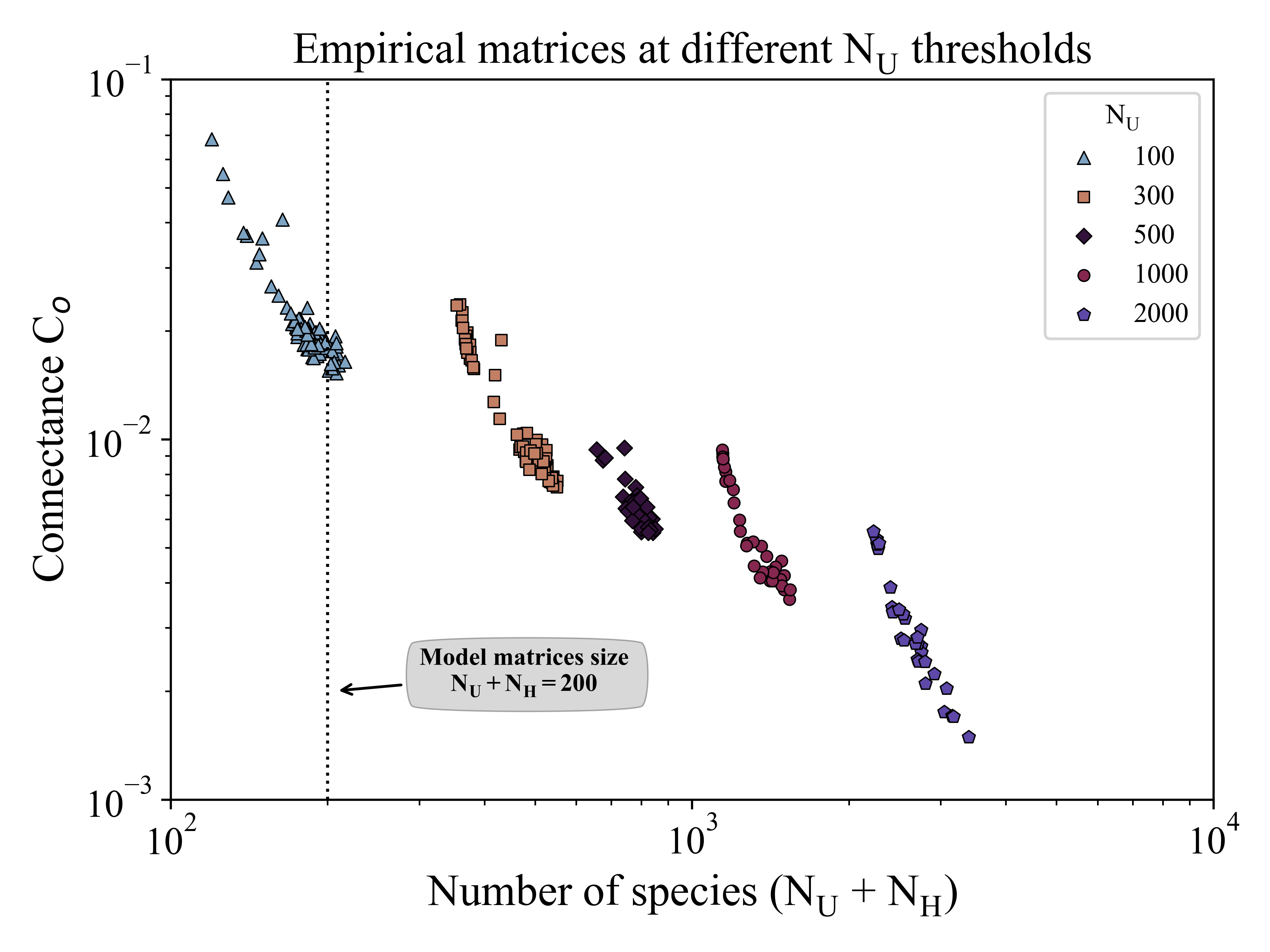} 
	\caption{Connectivity as a function of the number of species ($N_U+N_H$) for the empirical networks, at different $N_U$ thresholds. Note the log-log scale.}
	\label{fig_conn}        
\end{figure}

To avoid excessive computational costs, we consider small synthetic networks of $N_{U} = 100$ users and $N_{H} = 100$ hashtags with random connections across guilds, and density (connectance) $C_o \sim 10^{-2}$. We do so to match the same order of magnitude of empirical networks when we take $N_{U} = 100$, see blue triangles in Fig.~\ref{fig_conn}. We assign the same initial abundance $n_0=0.2$ to all the users and hashtags, and the same intrinsic growth rates $\rho_U = \rho_H = 1$. Results in the following subsections correspond to an average over 10 different realisations for each combination of  these parameters.

\subsubsection{Pre-event steady state}
In first place, we pay attention to the abundances of individual species along the simulation time, and check for regions on the parameter space where extinctions may occur. We consider that a species goes extinct if its abundance falls below $10^{-4}$.  

\begin{figure}[h!]
\centering
\def\stackalignment{l}
\topinset{\bfseries(a)}{\includegraphics[width=0.32\textwidth]{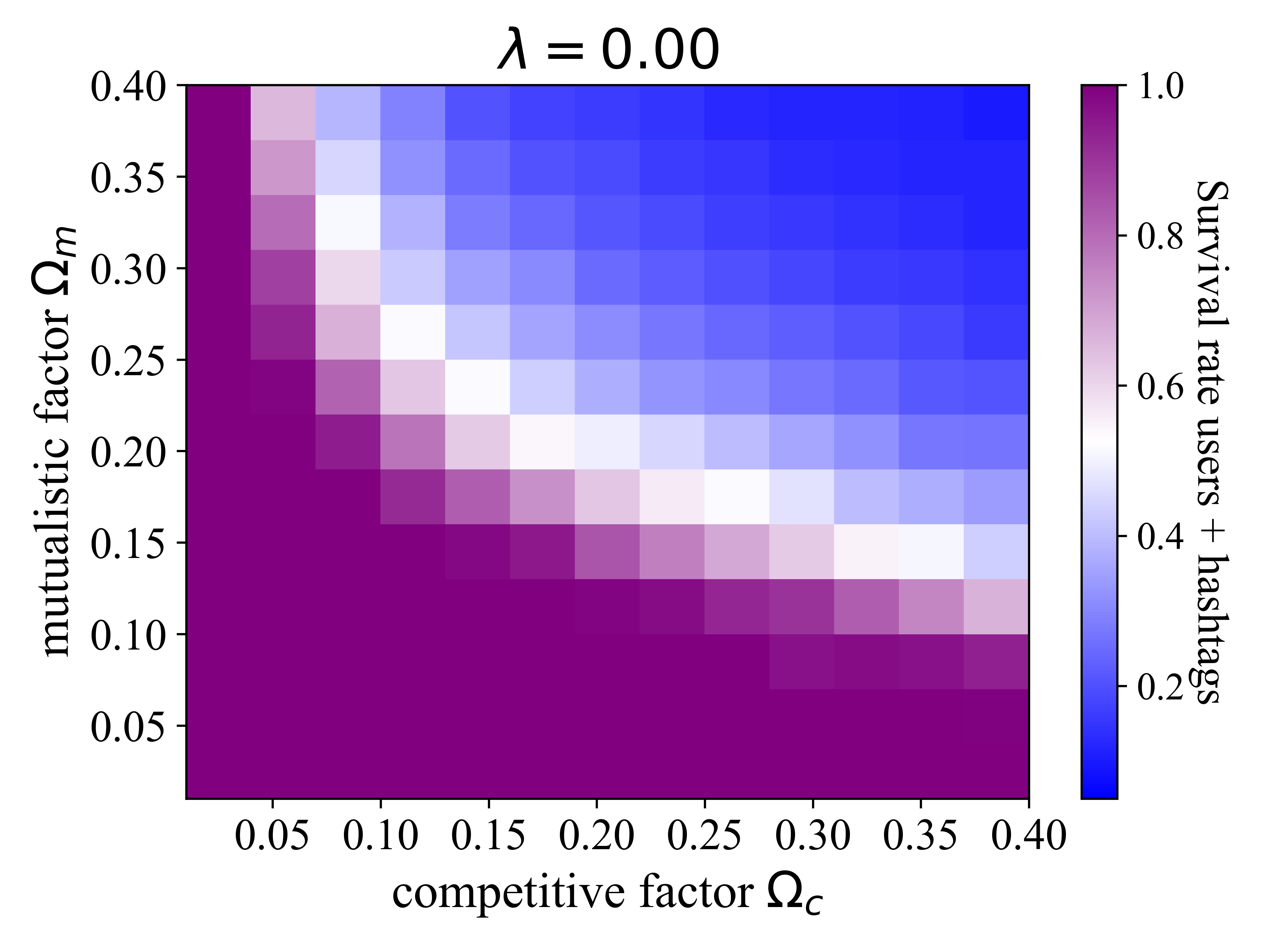}}{0.15in}{0.05in}
\topinset{\bfseries(b)}{\includegraphics[width=0.32\textwidth]{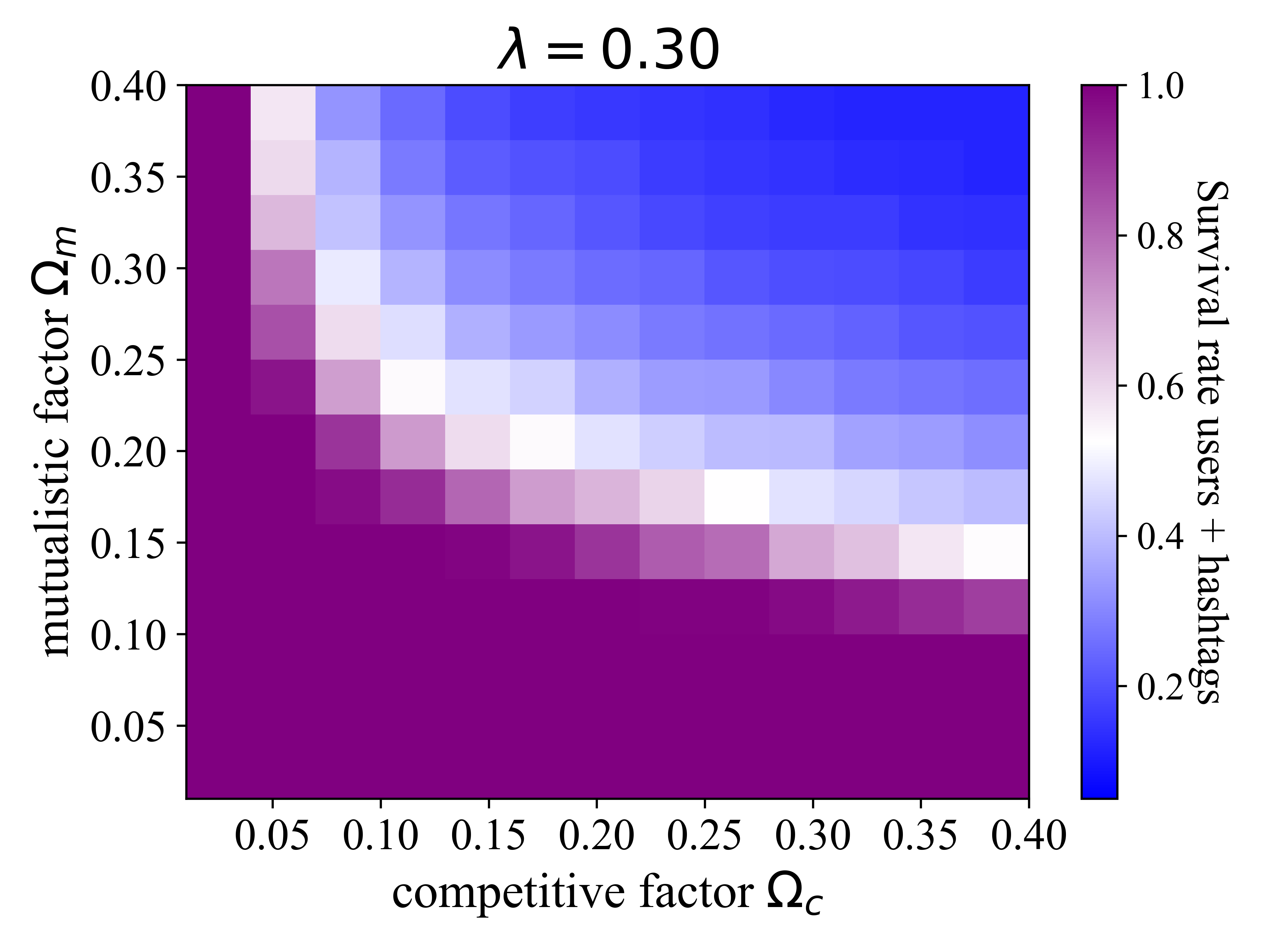}}{0.15in}{0.05in}
\topinset{\bfseries(c)}{\includegraphics[width=0.32\textwidth]{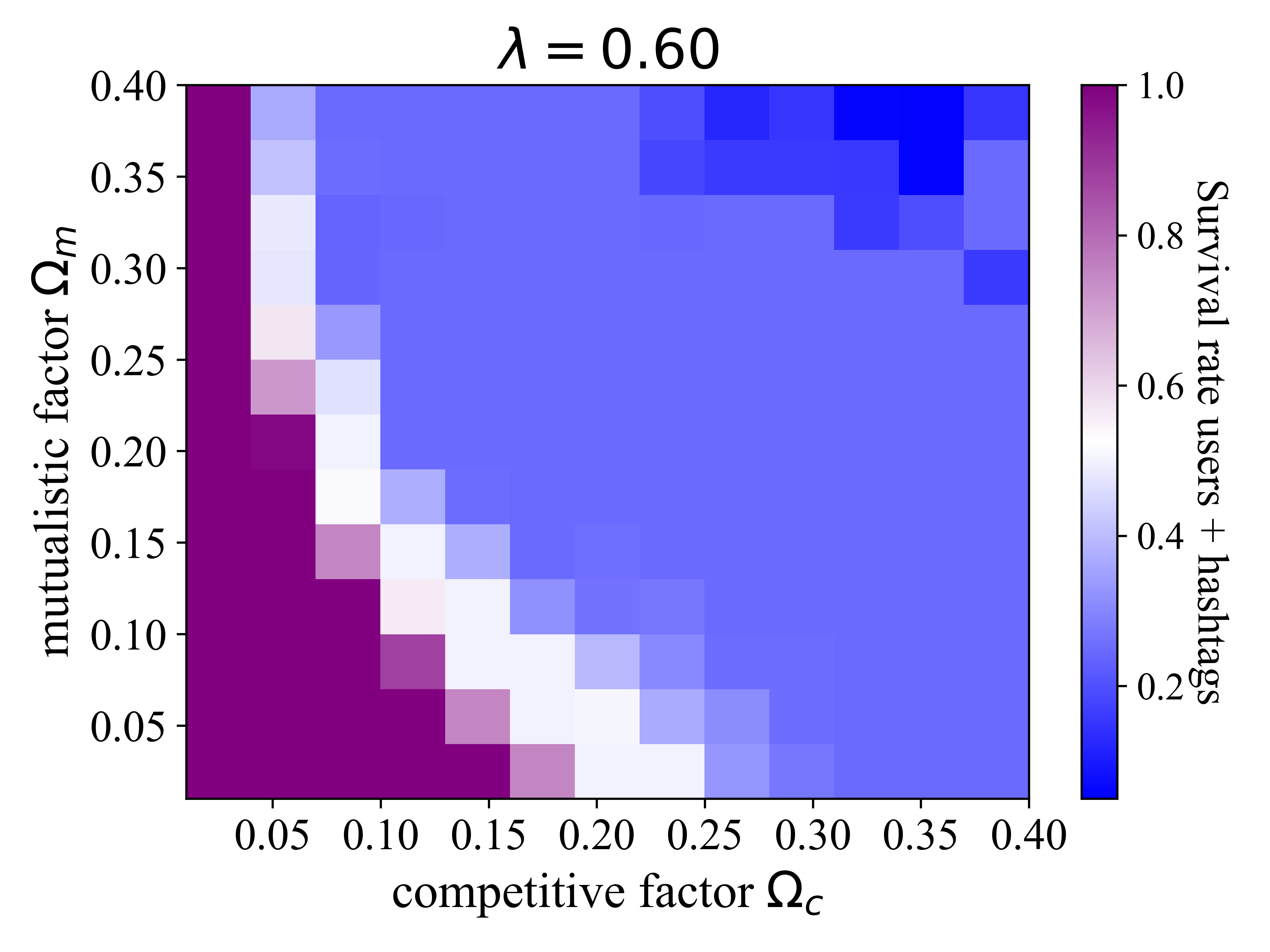}}{0.15in}{0.05in}
\vspace{-.1cm}\caption{\textbf{Survival rate at the pre-event steady state:} two-dimensional plots in the $\Omega_m- \Omega_c$ parameter space showing steady survival rate of the species for  different values of the inter-intra competition parameter. }\label{fig:survival_1}
\end{figure}

Fig. \ref{fig:survival_1} shows the fraction of especies survival in the two dimensional plot in the $\Omega_m- \Omega_c$ parameter space. For all the cases, we observe that as $\Omega_m$ and $\Omega_c$ increase, extinctions start to occur, even for favorable configurations of the system in which $\Omega_m> \Omega_c$.
 
As expected, for low values of the inter-intra competition parameter $\lambda=0$ and $\lambda=0.3$ the region in which extinction do not occur is wider. Under this configuration, the species compete more strongly within their topics, which correspond to just a fraction of all the species on the system.  On the contrary, as we increase $\lambda$, the region of extinctions may increase, since now each specie start to compete with a higher fraction of the system, making the system more susceptible to the values $\Omega_m, \Omega_c$. In order to guarentee the maximal survival species in the system prior to the introduction of the events, we will restrict our exploration on the $\Omega_m- \Omega_c$ to the interval $[0.01, 0.1]$. For the sake of simplicity, from now on we will show the results just for $\lambda=0.6$, which correspond to the case presented in the main text.

\begin{figure}[h!]
\centering
\def\stackalignment{l}
\topinset{\bfseries(a)}{\includegraphics[width=0.45\textwidth]{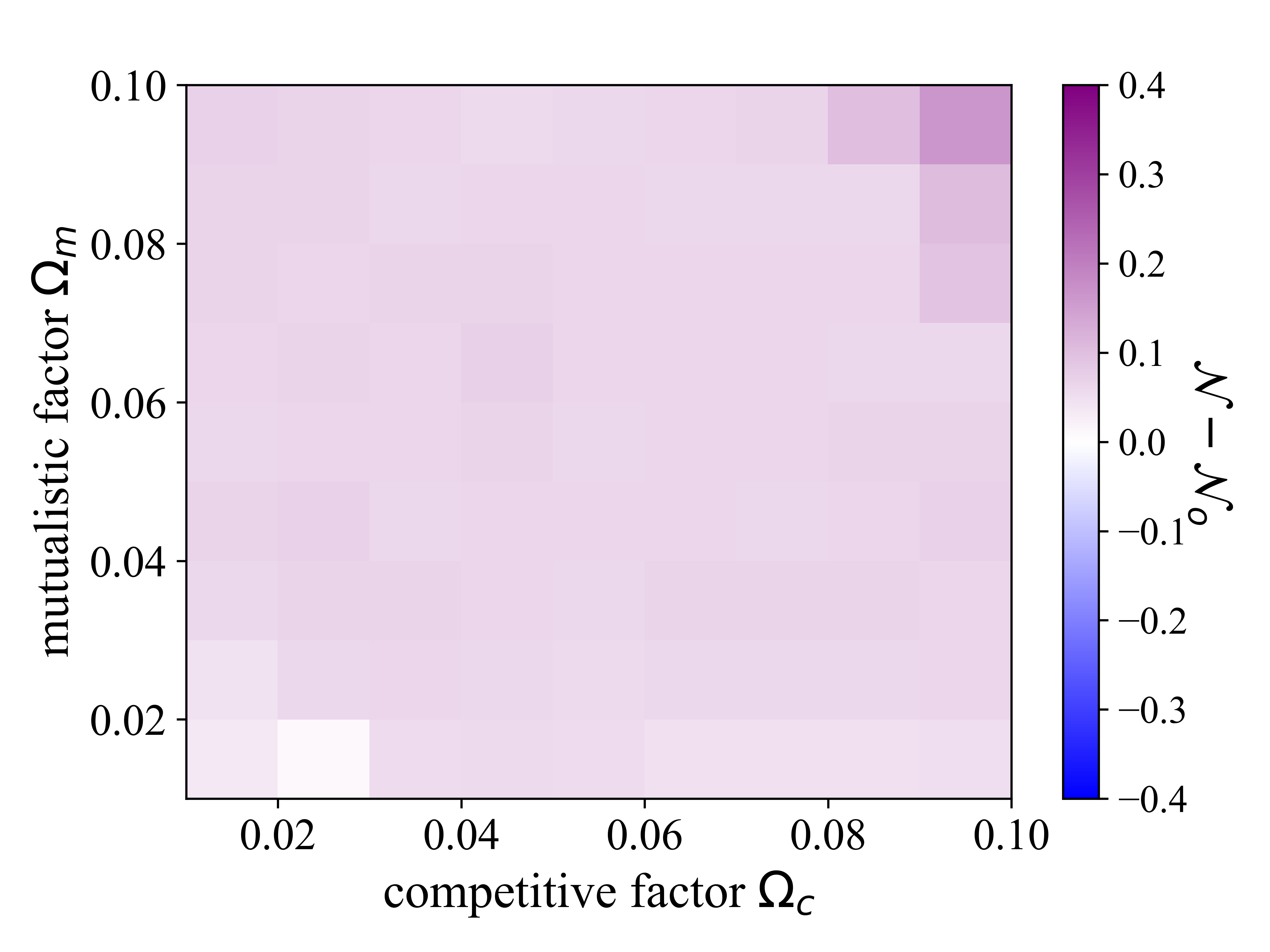}}{0.15in}{0.05in}
\topinset{\bfseries(b)}{\includegraphics[width=0.45\textwidth]{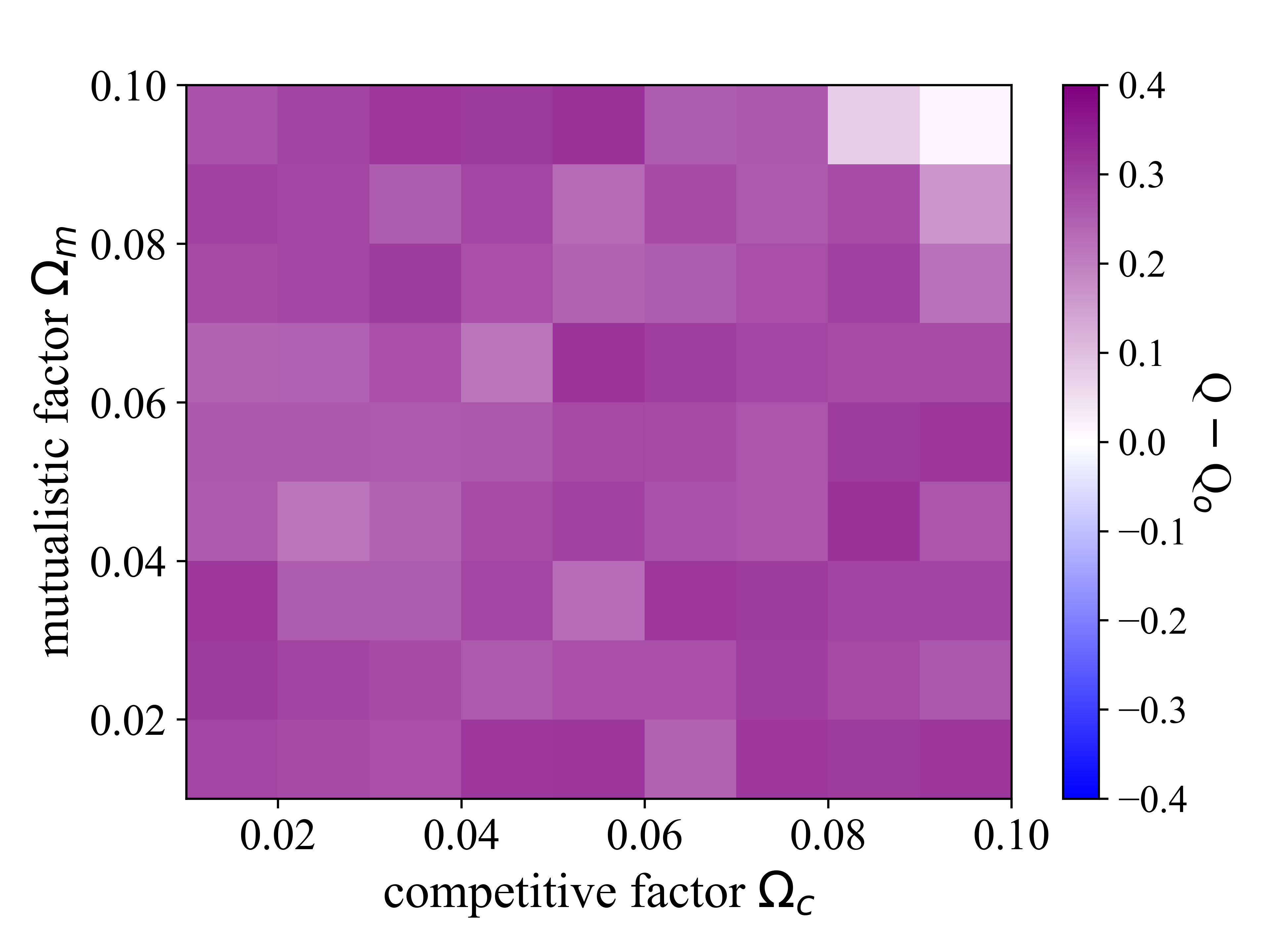}}{0.15in}{0.05in}
\vspace{-.1cm}\caption{\textbf{Structural measures at the pre-event steady state:} two-dimensional plots in the $\Omega_m- \Omega_c$ parameter space showing the evolution $\altmathcal{N}$ and Q before the external event  for  $\lambda=0.6$. $\altmathcal{N}_o$ and Q$_o$, correspond to the values at the beginning of the simulation.}\label{fig:metrics_1}
\end{figure}

Fig. \ref{fig:metrics_1} presents the two dimensional plot in the $\Omega_m- \Omega_c$ parameter space before the event . Each point within the plot corresponds to the values for each structural measure minus the values of $\altmathcal{N}_o$ and Q$_o$ measured  at the beginning of the simulation.  We found that the system becomes highly modular for a wide range of the $\Omega_m- \Omega_c$ values, while nestedness remain low, except for the small regions where species extinctions occur. These results are robust for  all the values of $\lambda$ that were considered.  Hence, by introducing species niche aligned to a certain number of topics, we were able to mimic the prescribed organisation in topical blocks observed in the empirical datasets. The modular architecture arises from the random one at the end of the optimization process. 

\subsubsection{Introducion of an external event}
In the current section, we present the results of the simulation after the introduction of an external event that correspond to a  shift in the users' niches. Since we observe an equivalent structural behavior after introducing different types of events, the following plots only show the results of the case of the sudden event described in section \ref{sud_ev}.

\begin{figure}[h!]
\centering
\def\stackalignment{l}
\topinset{\bfseries(a)}{\includegraphics[width=0.32\textwidth]{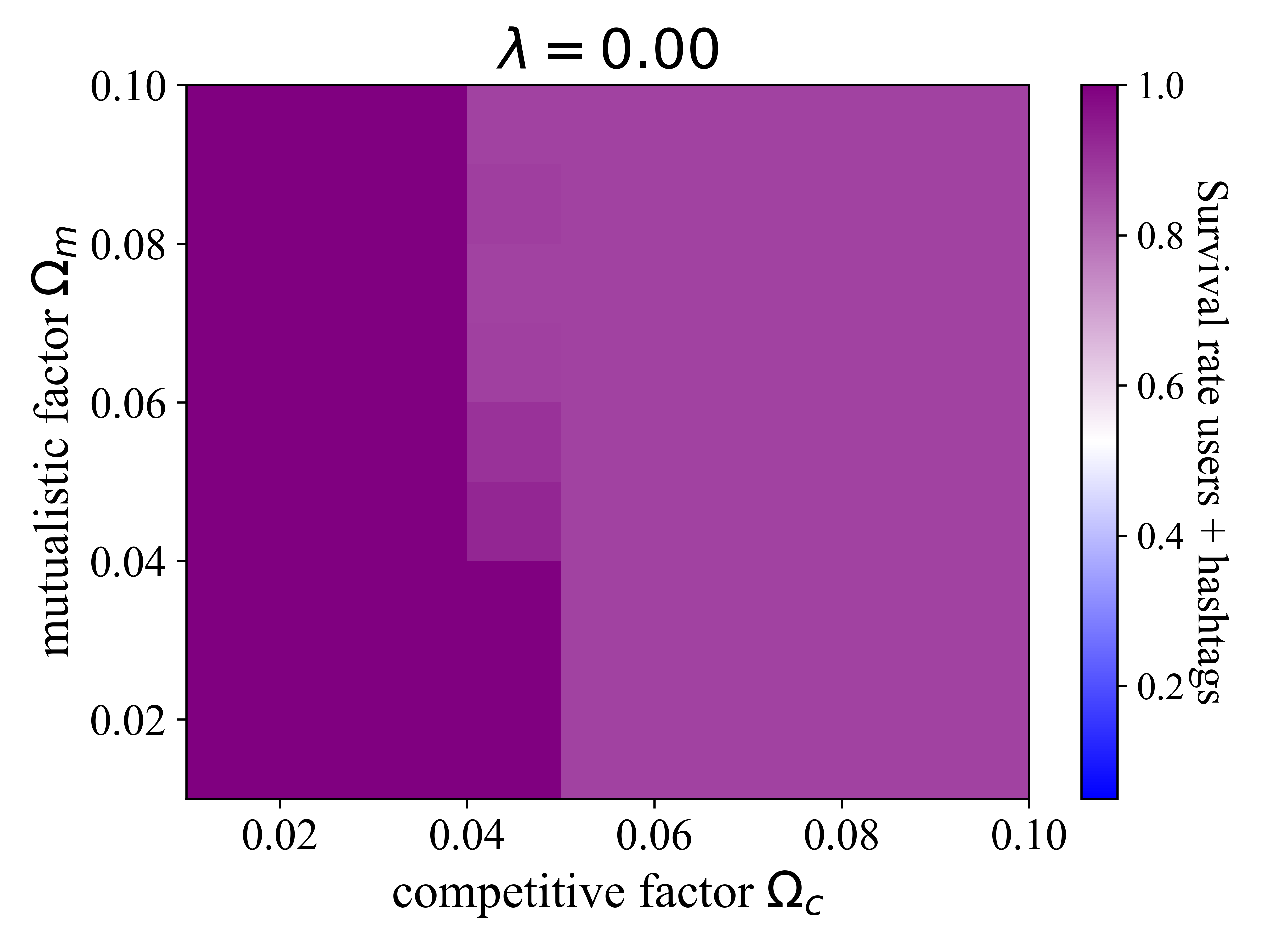}}{0.15in}{0.05in}
\topinset{\bfseries(b)}{\includegraphics[width=0.32\textwidth]{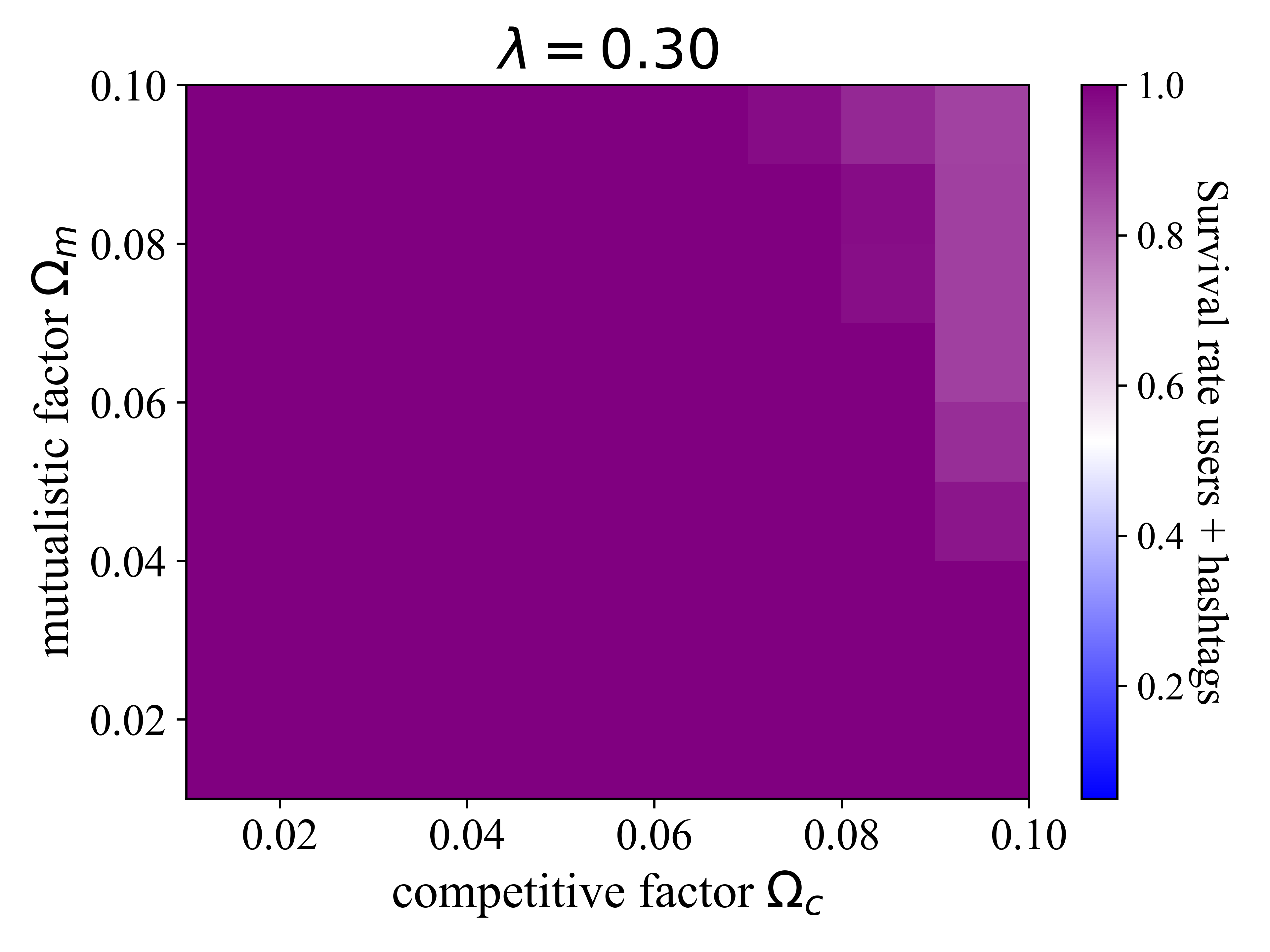}}{0.15in}{0.05in}
\topinset{\bfseries(c)}{\includegraphics[width=0.32\textwidth]{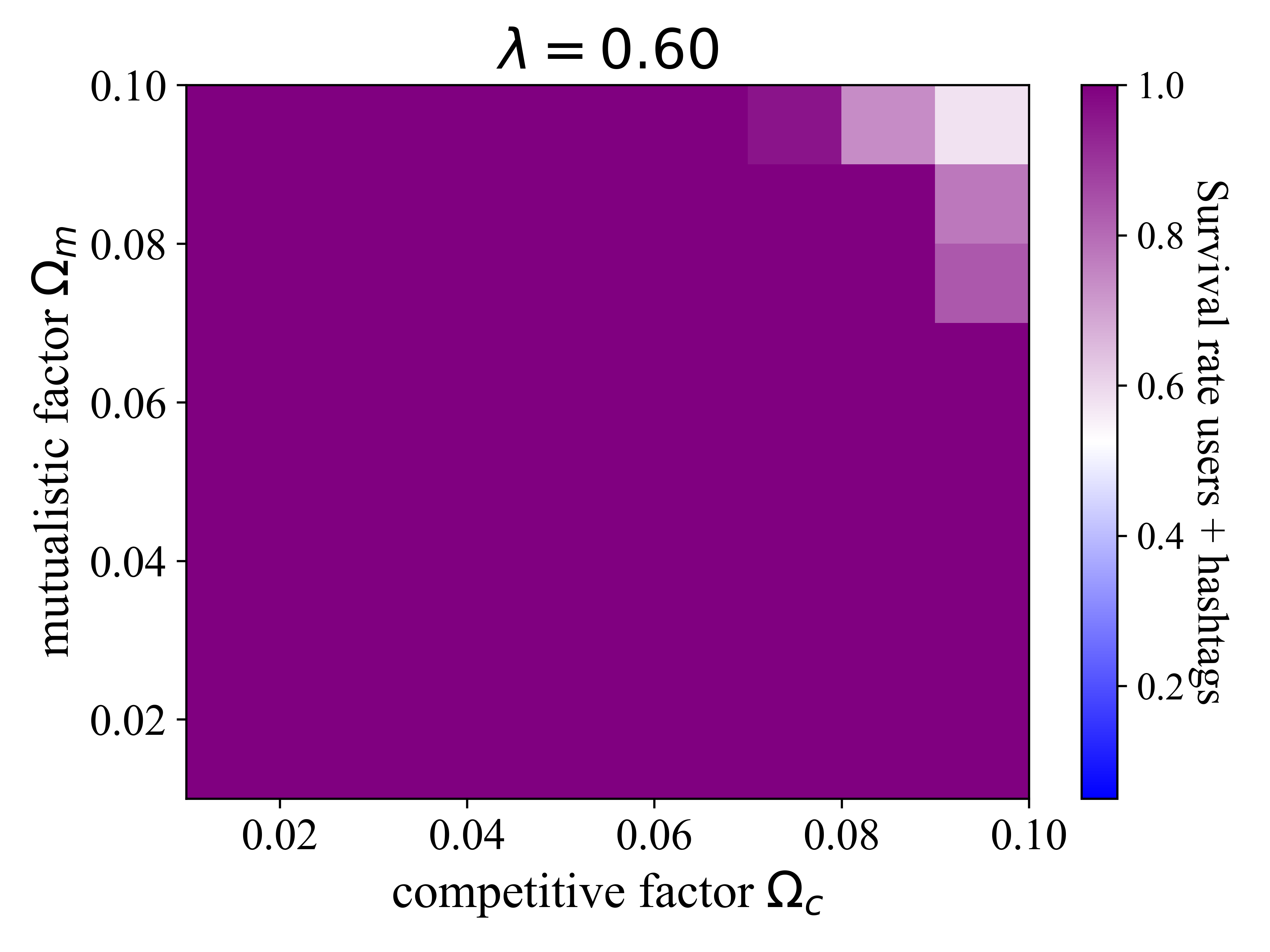}}{0.15in}{0.05in}

\vspace{-.1cm}\caption{\textbf{Survival rate at $t > t_{E}$:} two-dimensional plots in the $\Omega_m- \Omega_c$ parameter space showing survival rate of the species  at the end of the simulation, for different values of the inter-intra competition parameter $\lambda$. }\label{fig:survival_2}
\end{figure}

As is shown in Fig. \ref{fig:survival_2}, for $\lambda=0$ a considerable number of especies goes extinct by the end of the simulation time. This results is not surprising, since at the onset of the event the single topic configuration increases the competition among species. The $\lambda$ parameter helps to balance the intense competition between the species, therefore, we observe a decrese on the amount of extinctions as $\lambda$ goes higher.

\begin{figure}[h!]
\centering
\def\stackalignment{l}
\topinset{\bfseries(a)}{\includegraphics[width=0.45\textwidth]{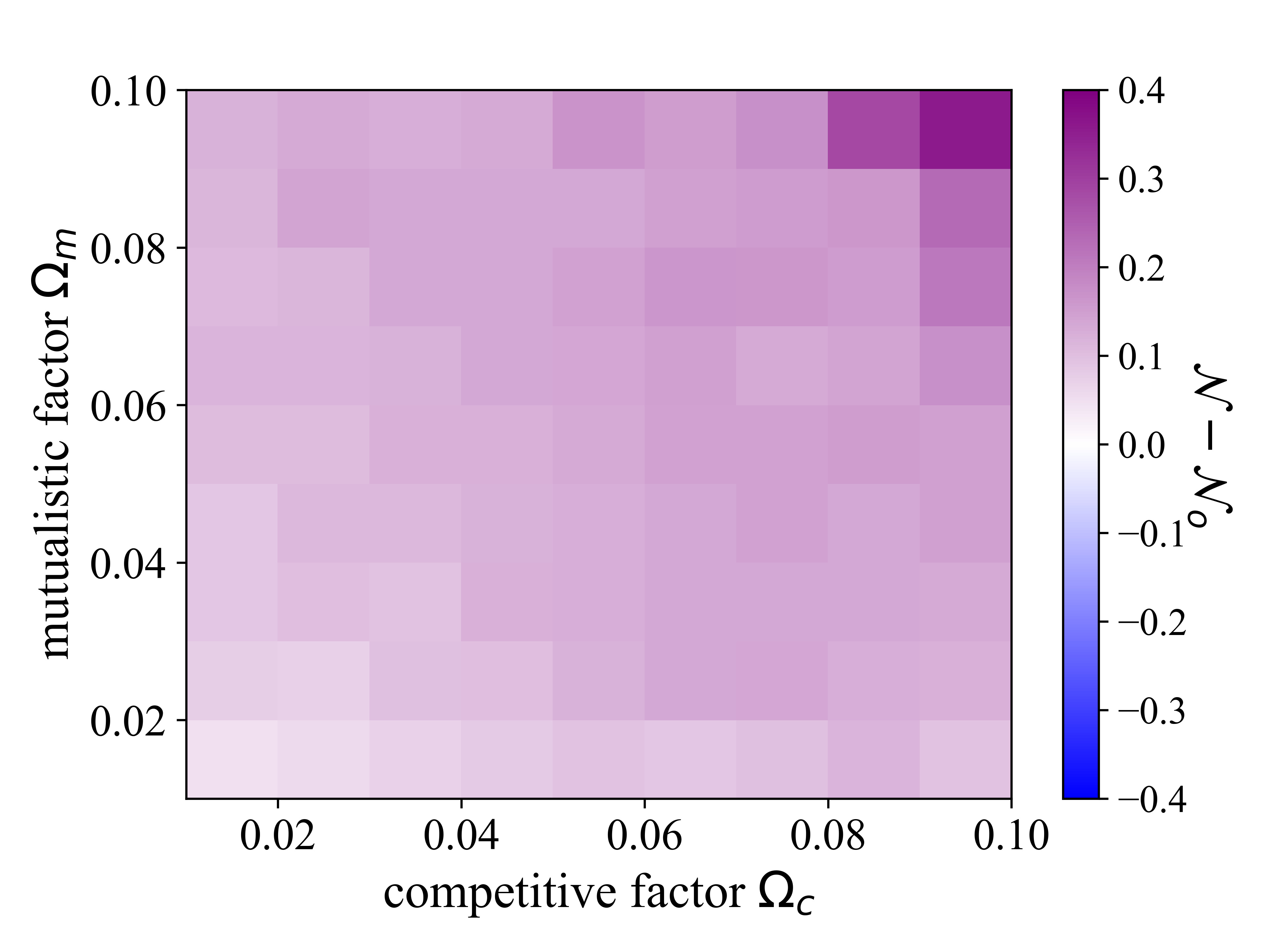}}{0.15in}{0.05in}
\topinset{\bfseries(b)}{\includegraphics[width=0.45\textwidth]{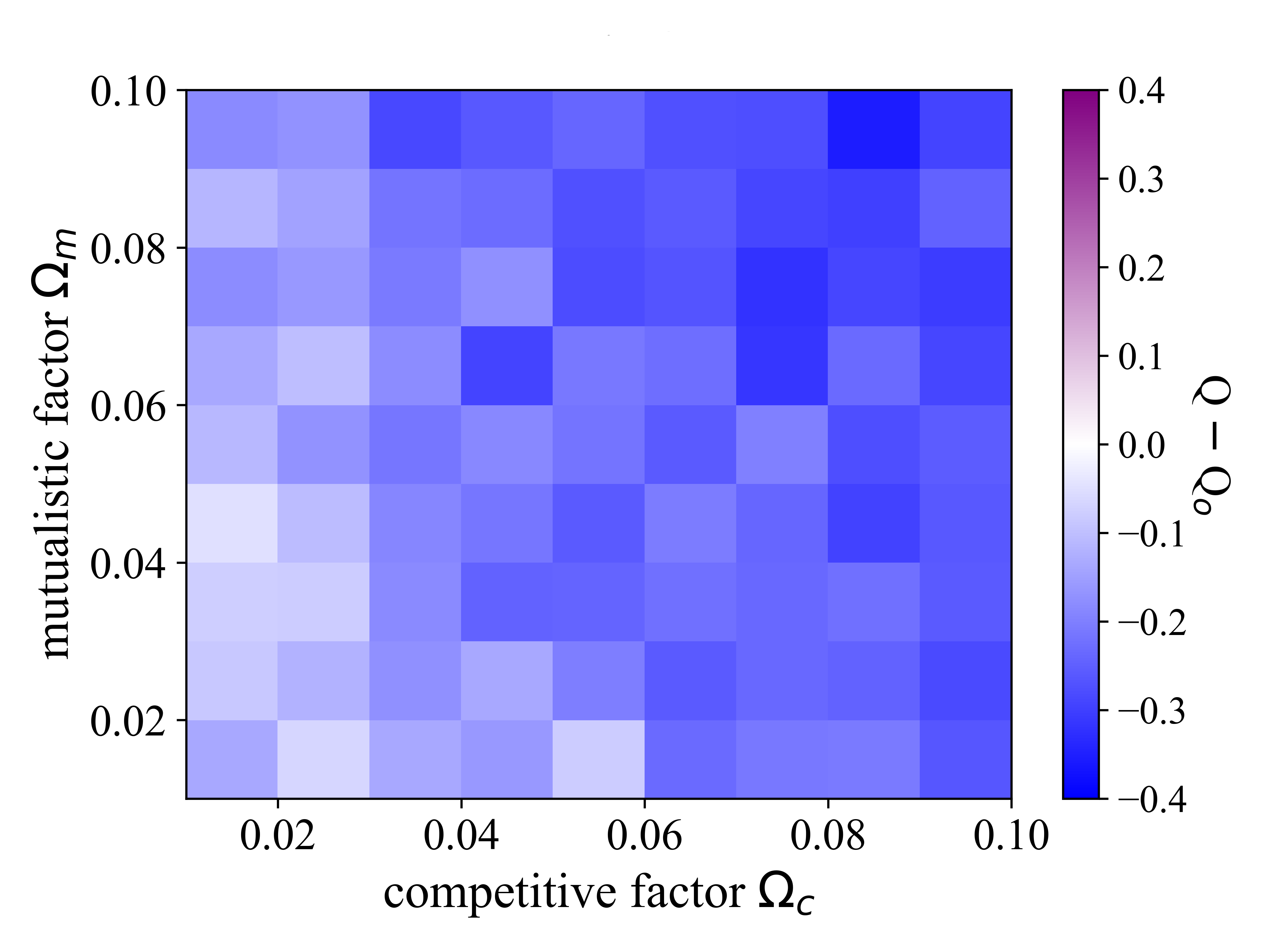}}{0.15in}{0.05in}
\vspace{-.1cm}\caption{\textbf{Structural measures at $t > t_{E}$:} Two-dimensional plots in the $\Omega_m- \Omega_c$ parameter space showing the evolution of $\altmathcal{N}$ and Q after the external event ($\lambda=0.6$). $\altmathcal{N}_{o}$ and Q$_{o}$, correspond to the values at the beginning of the simulation. }\label{fig:metrics_2}
\end{figure}

Turning our attention to the structural evolution of the system, we observe that this single topic configuration induced by the introduction of the event, provokes a structural transition from modular to a global nested pattern on the system, for a wide range of the $\Omega_m- \Omega_c$ parameters, see Fig \ref{fig:metrics_2}.  Again, in Fig \ref{fig:metrics_2}, $\altmathcal{N}_{o}$ and Q$_{o}$ correspond to the values for nestedness and modularity at the eginning of the simulation. We observe that modularity drops abruptly, while nestedness increases. This result is in accordance with our empirical observations reported above and in the maint text.

\subsubsection{Self-similar nested arrangements}
\begin{figure}[h!]
\centering
\def\stackalignment{l}
\topinset{\bfseries(a)}{\includegraphics[width=0.45\textwidth]{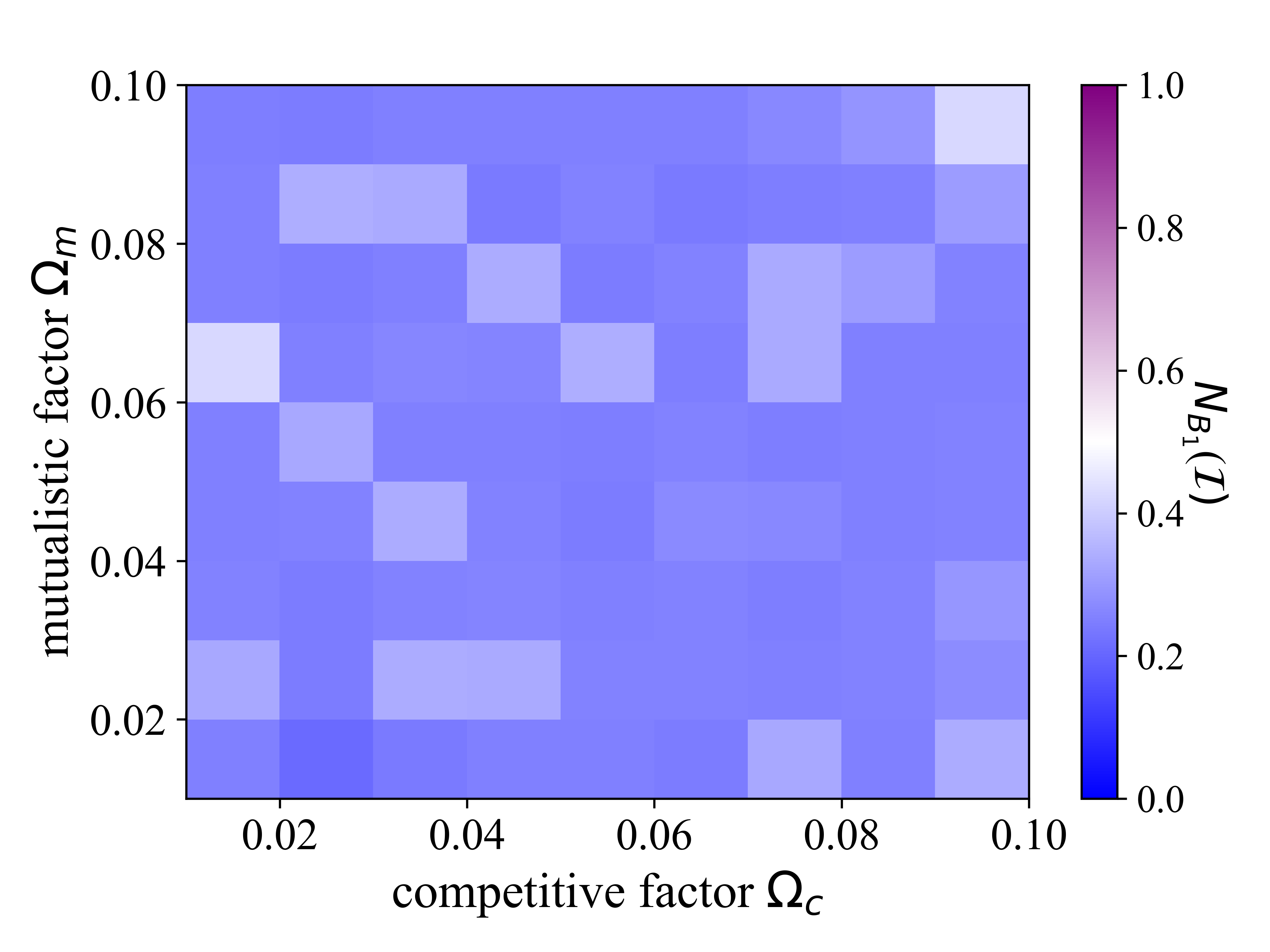}}{0.15in}{0.05in}
\topinset{\bfseries(b)}{\includegraphics[width=0.45\textwidth]{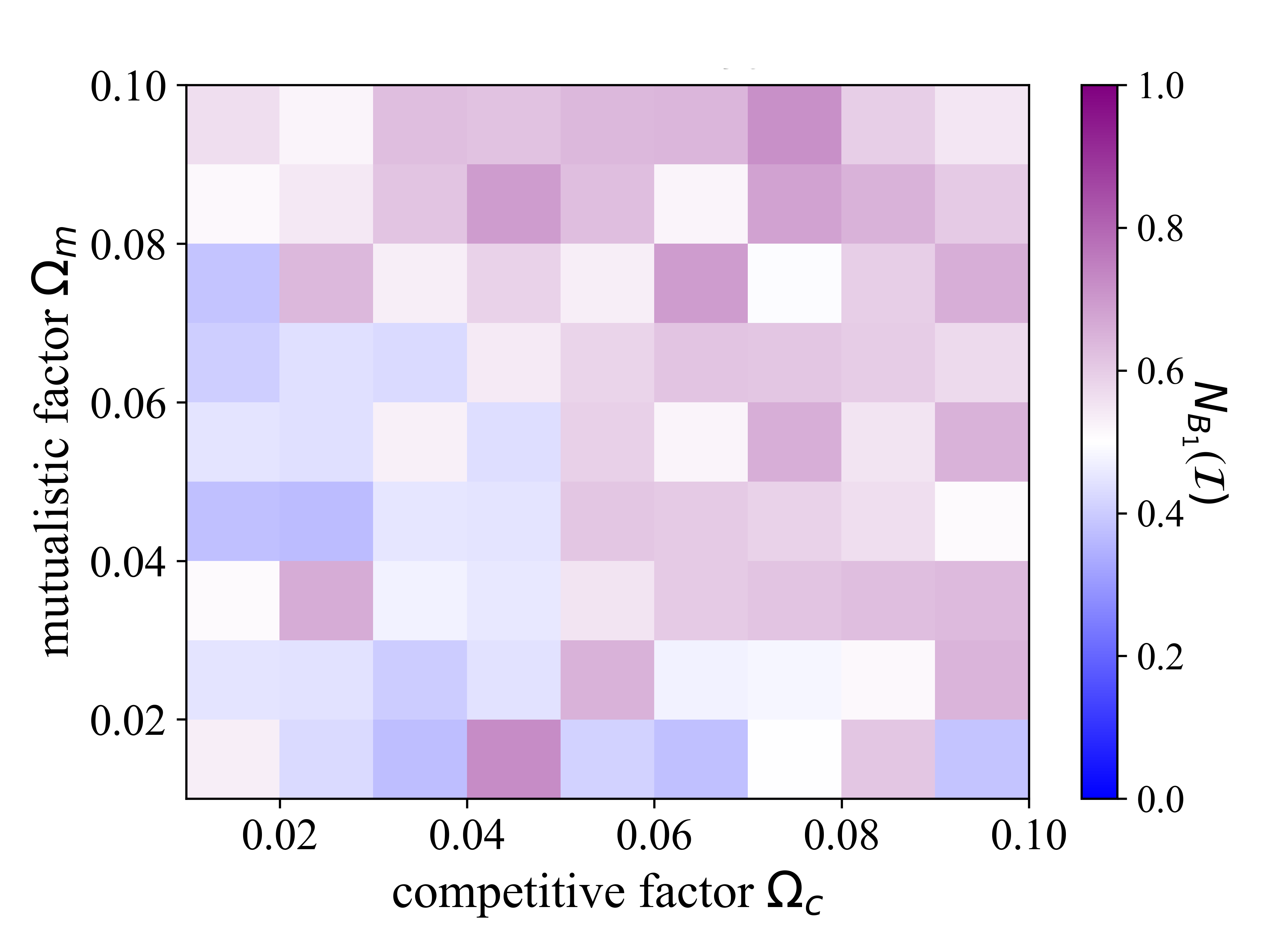}}{0.15in}{0.05in}

\vspace{-.1cm}\caption{\textbf{Relative size of the largest nested block $(N_{B_1} /N)$ before and after the external event:} two-dimensional plots in the $\Omega_m- \Omega_c$ parameter space showing the relative size of the largest nested block before (panel (a)) and after (panel (b)) the external event with $\lambda=0.6$.  }\label{fig:ibn}
\end{figure}

Finally, we also explored the structural evolution of the system by means of the in-block nestedness function $\altmathcal{I}$  \cite{sole2018revealing}. This exploration confirms the general character of the fluctuating nested self-similar organization. Figure~\ref{fig:ibn} shows the relative size of the largest nested blocks $N_{B_1}(\altmathcal{I})/N$, before (panel (a)) and after (panel (b)) the external event. Before the event, we observe that in general, for all the parameter space, the size of the largest nested block constitutes a $25\%$ of the whole network, approximately, i.e, the user are  evenly aligned over the four predefined topics. After the event, we observed how a state of global consensus is emerging, as $N_{B_1}/N$ increases over all the parameter space, representing more than $50\%$ of the size of the network in most of the cases.


\bibliographystyle{plain}
\bibliography{nest}